\documentclass[AMA,STIX1COL]{./WileyNJD-v2}

\usepackage{bm}
\usepackage[load=abbr]{siunitx} 
\usepackage{xr-hyper}
\usepackage{hyperref}
\usepackage{cleveref}
\usepackage{subcaption}
\usepackage{xcolor, colortbl}
\usepackage{graphicx}
\usepackage{float}
\usepackage{framed}

\articletype{Original Article}%
\begin{document}

\title{A scalable solver for a stochastic, hybrid cellular automaton model of personalized breast cancer therapy}

\author[1]{Xiaoran Lai$\dagger$}
\author[1]{H\aa kon A. Task\'en$\dagger$}
\author[2]{Torgeir Mo}
\author[3]{Simon W. Funke}
\author[1,4]{Arnoldo Frigessi}
\author[3]{Marie E. Rognes}
\author[1]{Alvaro K\"ohn-Luque*}

\authormark{X. Lai \textsc{et al}}

\address[1]{\orgdiv{Oslo Centre for Biostatistics and Epidemiology, Faculty of Medicine}, \orgname{ University of Oslo},\orgaddress{\state{} \country{Norway}}}
\address[2]{\orgdiv{Institute for Cancer Research}, \orgname{Oslo University Hospital}\orgaddress{\state{}, \country{Norway}}}
\address[3]{\orgdiv{Simula Research Laboratory}\orgname{}, \orgaddress{\state{Lysaker}, \country{Norway}}}
\address[4]{\orgdiv{Oslo Centre for Biostatistics and Epidemiology}, \orgname{Oslo University Hospital}\orgaddress{\state{}, \country{Norway}}}

\corres{* \email{alvaro.kohn-luque@medisin.uio.no}}
\footnotetext[1]{The two authors contributed  equally to this paper.}

\abstract[Summary]{Mathematical modeling and simulation is a promising approach to personalized cancer medicine. Yet, the complexity, heterogeneity and multi-scale nature of cancer pose severe computational challenges. 
  A powerful approach to mimic biological complexity and to describe the dynamical exchange of information across different scales, is to couple discrete cell-based models with continuous models using hybrid cellular automata. However, such models become computationally very expensive when considering clinically relevant cancer portions.
  While efficient approaches to parallelize continuous models exist, their coupling with discrete models, and in particular with cellular automata, calls for more elaborated solutions.
The model consists of multiple ordinary and partial differential equations coupled in space and time with stochastic cellular automaton rules. 
  Building upon FEniCS, a popular and powerful scientific computing platform for solving partial differential equations, we developed parallel algorithms to link stochastic cellular automata with differential equations (\url{https://bitbucket.org/HTasken/cansim}).
  The algorithms minimize the communication between processes sharing cellular automata neighborhood values and allow reproducibility during stochastic updates.
  We demonstrated the potential of our solution on a complex hybrid cellular automaton model of breast cancer under combination chemotherapy. 
  In performance tests, on a single-core processor, we obtained almost linear scaling with an increasing problem size, while weak parallel scaling showed moderate growth in solving time relative to increase in problem size. 
  We applied the algorithm to a problem that is 500 times larger than previous work.
  This allowed us to run personalized therapy simulations based on heterogeneous cell density and tumor perfusion conditions estimated from magnetic resonance imaging data on an unprecedented scale.}

\keywords{multi-scale modeling, cancer modeling, parallel computing, domain decomposition, FEniCS, personalized cancer therapy, breast cancer}

\maketitle

\section{Introduction}

Mathematical modelling and computer simulations, informed by patient-specific clinical data, can be used to make personalized predictions of response to cancer therapy~\cite{rockne2019}. 
The methodology can in principle be used to design patient-specific treatment plans and constitutes a promising approach to personalised cancer medicine. 
One of the main goal is the development of quantitative and computational tools that can effectively and efficiently simulate the consequence of multiple therapeutical strategies in each patient. 
However the complexity, heterogeneity and multi-scale nature of cancer present severe computational challenges to that goal. 
Specifically, treated cancer tissue entails multiple interacting processes occurring at different spatio-temporal scales (for instance, intracellular signaling pathways, drug pharmacokinetics or single cell decisions such as division or death).
One approach to model the interactions is the hybrid cellular automata (HCA) framework coupling discrete cellular automata and continuous model components accounting for the different phenomena and scales~\cite{ribba2004, alarcon2005, gerlee2007, powathil2012}. 

To simulate HCA models, computational algorithms have to balance the numerical schemes used to efficiently solve the different discrete and continuous formalism together with the sharing of partial states between them~\cite{cilfone2015}.
Time can be easily wasted since some solvers are on hold while others need to reach stationary states for example. Moreover, as different solvers can have different numerical meshes in space and time, the sharing of states is generally not trivial. Additional sources of complexity to simulate cancer tissue are the large number of cells and the large heterogeneity present within the tissue. For instance, tumors often have areas with different cell densities and perfusion characteristics, cells and vessels could have different features, etc. As such heterogeneity is known to impact treatment outcome, there is a need for large scale simulations that can capture it. Furthermore, a simulation algorithm will be useful in a clinical setting only if the computational time to run it is compatible with clinical decision making. Parallel computing is therefore expected to play a key role.

FEniCS is a finite element computing platform for solving partial differential equations (PDEs) that has been fundamentally designed for parallel processing~\cite{alnaes2015}. 
While FEniCS has already been extended to couple PDEs with ordinary differential equations (ODEs) associated with mesh nodes, the coupling with stochastic cellular automata (CA) models, required to solve HCA, is not straightforward. Specifically, mesh partitioning methods used by FEniCS are not suitable for applying CA rules depending on values in different CA meshes. We solved this problem with an algorithm that sets up a map over what information needs to be exchanged between processes in each CA update. Additionally, we implement a priority system to resolve possible conflicts of multiple processes attempting to set a CA node value simultaneously. This, together with an efficient and light implementation of random number generation, allows us to reproduce stochastic model simulations, produce unbiased results and claim statistical significance.

To demonstrate the potential of our approach and algorithms, we consider an updated multi-scale HCA model, previously used for simulating personalised breast cancer therapy~\cite{lai2019, laiphd2019}. The model reproduces and explains treatment outcomes at the level of individuals. For tumors that did not respond to therapy, model simulations were used to suggest more successful regimes, depending on the patient's individual characteristics. These results were promising but were limited to small sections of the tumor with only a few hundred biological cells, and simulated on a single CPU. In order to simulate clinically relevant pieces of tumors and to capture the tumor heterogeneity as observed in magnetic resonance imaging (MRI) data, an efficient and scalable parallel solver was developed, and presented here in this paper. We link the stochastic CA model and the continuous PDE and ODE models to run efficiently and in parallel. In addition to testing the scalability of single core performance, we perform a weak scaling study on cluster up to 80 cores with increasing problem size. Finally, we show that it is now possible to run simulations of 2D tumor sections that are approximately 500 times larger than previous work~\cite{lai2019, laiphd2019}.

\section{Methods}
\label{sec:methods}

\subsection{Magnetic resonance imaging (MRI)}
\label{subsec:MRI}
Variation in cellular and vascular density across breast tumor tissue of one patient was assessed with magnetic resonance imaging (MRI). The patient underwent MRI examinations before the start of treatment, and after 1  and 12 weeks of neoadjuvant treatment. Examinations were performed on an ESPREE $\SI{1.5}{\tesla}$ MR scanner (Siemens, Erlangen, Germany) equipped with a phased-array bilateral breast coil (CP breast coil, Siemens, Erlangen, Germany). The MRI protocol was a state-of-the-art MRI protocol \cite{Mann2015} with T2-weighted, diffusion-weighted (DW), and dynamic contrast-enhanced (DCE) MRI. DCE-images were acquired using  a radial, spoiled gradient echo with k-space weighted image contrast (KWIC), using spectral adiabatic inversion recovery (SPAIR) for fat-suppression (TE = $\SI{2.59}{\milli\second}$, TR = $\SI{5.46}{\milli\second}$, flip angle = $15^{\circ}$, field of view = $\SI{320}{\mm} \times \SI{320}{\mm}$, in-plane resolution = $\SI{1}{\mm} \times \SI{1}{\mm}$, slice thickness $= \SI{15}{\mm}$). After eight pre-contrast series, the contrast agent (Gadovist, Bayer Pharma, Germany) was administered at a dosage of $\SI{0.8}{\mmol\per\kg}$, at a rate of $\SI{3}{\ml\per\second}$, followed by a $\SI{20}{\ml}$ saline flush. Subsequently, 32 post contrast image-series were acquired at a frequency of $\SI{1/13}{\per\second}$.  

DCE images were analysed using an extended Tofts two-compartment pharmacokinetic model, yielding $\SI{1}{\mm} \times \SI{1}{\mm}$-resolution maps of \emph{in-vivo} perfusion parameters in the tumor. Perfusion parameters calculated includes the permeability-surface area product, $K_{trans}$ of the vasculature, volume fraction of the vascular space, $v_p$ and volume fraction of the extravascular, extracellular space, $v_e$. Additionally, the cellular density was estimated as $v_c = 1-(v_p + v_e)$. Image analysis was performed using nICE (Nordic NeuroLab, Bergen, Norway). Patient specific arterial input function was sampled from a region of the right atrium of the heart, visible in the DCE images.

\subsection{Multi-scale mathematical model}
\label{subsec:math_model}
\begin{figure}[htp] 
   \begin{center} 
     \includegraphics[draft=false,width=\textwidth]{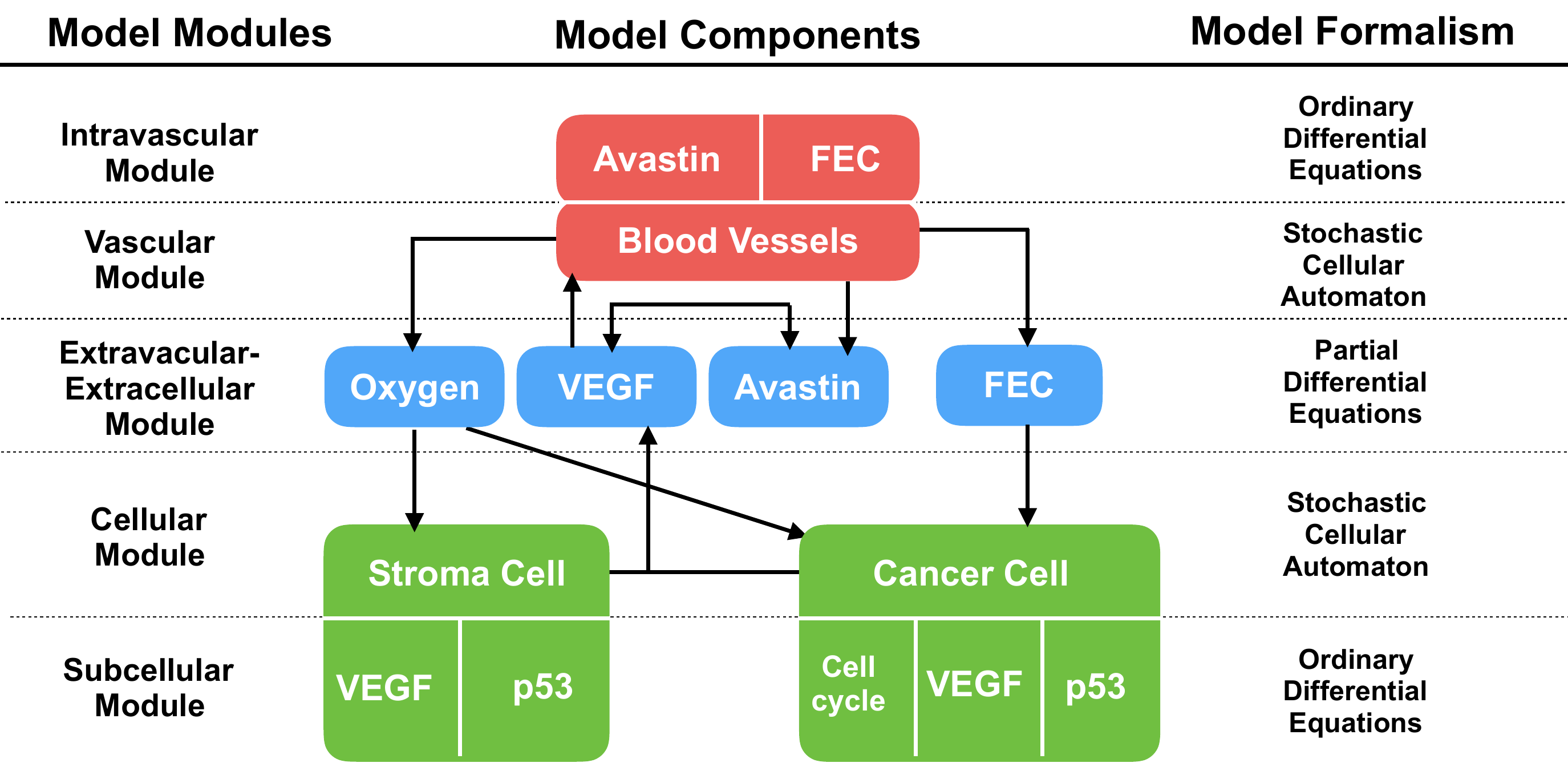}
   \end{center}
   \caption{Modular structure of the hybrid cellular automaton model for vascular tumour growth and combination therapy. The directions of arrows indicate influence between different components of each module. The last column shows the mathematical formalism used to describe each of the modules. }
   \label{fig:flowchart}
 \end{figure}

We begin by summarizing the multi-scale mathematical model of breast cancer and its treatment by a combination of injected drugs. We refer to our previous work for a further biological and clinical discussion of the model~\cite{lai2019}. The model accounts for the response of a 2D cross section of tumor tissue to a combination of chemotherapeutic and anti-angiogenic agents using hybrid cellular automata~\cite{ribba2004, alarcon2005}. Thus, a tumor section is represented by a finite regular square lattice $L$, consisting of a set of nodes labeled by their positions $\bm{x} \in L$, $\bm{x} = (i \Delta x,j \Delta x)$, $i=0,..,n-1, j=0,...,m-1$,  $\Delta x$ being the distance between nearest nodes. Biological cells and cross-sectional cuts of functional blood vessels are modeled as individual agents occupying a single lattice node. Microenvironmental factors in the tissue section, such as oxygen, are modeled as continuous variables over the domain $D=[0, (n-1)\Delta x] \times [0, (m-1)\Delta y] \subset \mathbb{R}^2$. Intracellular and intravascular processes are modeled as continuous variables associated to each cell and blood vessel, respectively. To account for cell and blood vessel dynamics and the molecular factors that control them, we build five interlinked model modules: the cellular, subcellular, vascular, intravascular and extravascular-extracellular modules. Figure~\ref{fig:flowchart} shows a diagram with the main components of each module, the interactions between them and the model formalism used in each case. A more detailed description of each module is provided below.
 
 \subsubsection{Stochastic cellular automaton for the cellular module}
\label{subsec:CA}
The presence of cancer and other cells, referred here as stroma, on the lattice sites at time $t$, is described by a function $\mathcal{L}$ that takes three possible values: $\mathcal{L}(\bm{x},t)=0$, if the site $\bm{x}$ is empty; $\mathcal{L}(\bm{x},t)=1$, if $\bm{x}$ is occupied by a cancer cell; or $\mathcal{L}(\bm{x},t)=2$, if $\bm{x}$ is ocuppied by a stroma cell.
At most one cell can occupy one site of the grid.
It is assumed that cancer cells can divide or die but cell movements are neglected. Cancer cell division is controlled by an internal cell cycle clock described in the subcellular module below. When a cell at position $\bm{x}_{\mathrm{cell}}$ is committed to divide, it can be killed by the chemotherapy agent $i=1,2,3$ with a probability given by:
\begin{equation}
\mathrm{I}\left(\frac{G_{i}(\bm{x}_{\mathrm{cell}}, t)}{\mathrm{max}(G_{i,0})}; 1, b_i\right),
\end{equation}
where $I(x;a,b)= \tfrac{\Gamma (a + b) }{\Gamma (a) + \Gamma (b)} \int_{0}^{x} y^{a-1} (1-y)^{b-1}\, \mathrm{d}y  $ is the regularized incomplete beta function. The value $\mathrm{max}(G_{i,0})$ refers to the maximal concentration of chemotherapy $i$ in the blood and $b_i$ is a parameter describing the sensitivity towards chemotherapy $i$. If the cell at position $\bm{x}$ dies at time $t$, it is removed from the computational grid by setting $L(\bm{x}, t) = 0$. If chemotherapy does not kill the proliferating cell, a daughter cell is placed at an empty (Moore) neighbor location with the highest oxygen pressure. If, however, no free space is available in the neighborhood, the cell cycle of the parent cell is reset to zero and no daughter cell is produced. 

Stroma cells in our cellular automaton do not proliferate or die. Thus, their role in the model is restricted to competition for space and oxygen with cancer cells and to production of the angiogenic factor VEGF under hypoxia (low oxygen tension).

\subsubsection{Stochastic cellular automaton for the vascular module}

We consider only perfused functional blood vessels and assume that all are cylindrical and perpendicular to the modelled tumor section.
The presence of cross-sectional vessel cuts in the lattice at time $t$ is given by the function $\mathcal{G}$, with $\mathcal{G}(\bm{x}, t) = 1$ representing the presence of a functional vessel and $\mathcal{G}(\bm{x},t) = 0$ its absence. 
Vessel dynamics is modeled by a birth-death process, with the probability of creating and removing vessels depending on the extracellular spatial distribution of the angiogenic factor VEGF. Specifically, we assume there is a range of VEGF concentrations where vessels are viable but outside that concentration range vessels are more likely to regress and disappear. Therefore,  at every fixed time step $\Delta v$, we define the probability of creating and removing a vessel at $\bm{x} \in L$ as:
\begin{align}
  P(\mathcal{G}(\bm{x},t + \Delta v) &= 0~|~\mathcal{G}(\bm{x},t) = 1)=p_{\text{death}}~~\text{if}~~~~~ V_{\bm{x}} <  \text{V}_{\text{low}}~~\text{or}~~ V_{\bm{x}} > \text{V}_{\text{high}}
  \notag\\
  P(\mathcal{G}(\bm{x},t + \Delta v ) &= 1~|~\mathcal{G}(\bm{x},t) = 0)=p_{\text{birth}}
  (\bm{x}, t) 
  ~~\text{if}~~~~~  \text{V}_{\text{low}} \leq V_{\bm{x}} \leq \text{V}_{\text{high}} 
  \label{eq:vascular_module}
\end{align}
where $\text{V}_{\text{low}}$ and  $\text{V}_{\text{high}}$ are lower and upper thresholds of VEGF concentration where functional vessels are viable, and \mbox{$V_{\bm{x}} = V(\bm{x},t)$} denotes the extracellular concentration of VEGF at location $\bm{x}$ at time $t$, which is described below. Extending our original model~\cite{lai2019}, we now assume that the probability of birth $p_{\textit{birth}}(\bm{x}, t)$ is proportional to the VEGF concentration $V(\bm{x},t)$, instead of being constant. Following Owen et al.~\cite{owen2011}, we also assume that at each time-step $\Delta v$, the expected value of new created vessels is given by $n_{\mathrm{new}}$:
\begin{align}
E[n_{\mathrm{new}}] = Pr_{\mathrm{sprout}} 2 \pi r_{0} h_{\mathrm{c}} \Delta v \int \frac{r_{\bm{x}}}{r_{0}}\frac{V_{\bm{x}}}{V_{\mathrm{sprout}} + V_{\bm{x}}}\sum_{\bm{x}\in \bm{ X_\mathrm{v}}} \delta(x - \bm{x}, t)\, \mathrm{d}x,
\end{align}
where $Pr_{\mathrm{sprout}}$ is the maximum probability that an endothelial sprout emerges from a surface of a vessel and forms a new vessel, $X_{\mathrm{v}} = \{\bm{x}:\mathcal{G}(x,t) = 1\}$ is the set of all blood vessel locations at time $t$. $r_{\bm{x}} = r(\bm{x}, t)$ is the vessel radius, $r_{0}$ is the average radius of initial vessels, $h_{\mathrm{c}}$ is the height of the simulated tissue layer and $V_{\mathrm{sprout}}$ is the VEGF concentration at which the probability is half-maximal.

\subsubsection{Systems of ordinary differential equations for the intravascular module}
The time-dependent concentrations of four drugs in the blood, i.e.~Avastin and a cocktail of three chemotherapies (Fluorouracil, Epirubicin, and Cyclophosphamide) together known as FEC100, are modelled by their respective pharmacokinetic equations, dosage and drug administration schedule. It is assumed that all vessels share the same drug concentration at a given time point. For Avastin, a two-compartment model is used:
\begin{subequations}
\begin{align}
\frac{dA_1}{dt} &= - \frac{q}{v_1}A_1 - \frac{cl}{v_1}A_1 + \frac{q}{v_2}A_2, \\
\frac{dA_2}{dt} &= - \frac{q}{v_2}A_2 - \frac{cl}{v_1}A_1, 
\end{align}
\label{eq:intraVEGFinhibitor}%
\end{subequations}
where $A_1=A_1(t)$ and $A_2=A_2(t)$ are the concentrations of Avastin at time $t$ in the plasma and peripheral compartments, respectively, and $q$, $cl$, $v_1$, $v_2$, are kinetic and compartment volume parameters. For Fluorouracil a single compartment model is used:
\begin{align}
\frac{d G_{1,0}}{d t} &= - \frac{v_{max}}{k_m + G_{1,0}} G_{1,0}, \quad t \neq t_{D_n}\\
\intertext{ and }
 G_{1,0} (t_{d_n}) &= \frac{d_n}{V}, \quad n = 1 \ldots N
\end{align}
where $G_{1,0}=G_{1,0}(t)$ is the plasma concentration at time $t$, $d_n$ is the $n$-th dose, and $v_{max}$, $k_m$ and $V$ are kinetic parameters. 
For Epirubicin, a three-compartment model is used:
\begin{subequations}
\begin{align}
\frac{dG_{2,0}}{dt} &= - \frac{q_2}{w_1}G_{2,1} - \frac{q_3}{w_1}G_{2,2} - \frac{cl_2}{w_1}G_{2,0} + \frac{q_3}{w_3}G_{2,2} + \frac{q_2}{w_2}G_{2,1}, \\
\frac{dG_{2,1}}{dt} &= - \frac{q_2}{w_2}G_{2,1} + \frac{q_2}{w_1}G_{2,0}, \\
\frac{dG_{2,2}}{dt} &= - \frac{q_3}{w_3}G_{2,2} + \frac{q_3}{w_1}G_{2,0},
\intertext{ and }
 G_{2,0} (t_{d_n}) &= \frac{d_n}{V_1}, \quad n = 1 \ldots N
\end{align}
\end{subequations}
where $G_{2,0}=G_{2,0}(t)$ is the plasma concentration at time $t$ and $G_{2,1}=G_{2,1}(t)$ and $G_{2,2}=G_{2,2}(t)$ are concentrations in two peripheral compartments at time $t$. The parameters $q_1$, $q_2$, $q_3$, $cl_2$ are kinetic rates and $w_1$, $w_2$, $w_3$ are compartment volumes. For Cyclophosphamide, a single compartment model is again used:
\begin{align}
\frac{dG_{3,0}}{dt} &= -\frac{cl_3}{u} G_{3,0}, 
\intertext{ and }
G_{3,0} (t_{d_n}) &= \frac{d_n}{V}, \quad n = 1 \ldots N
\label{eq:intrachemo3}
\end{align}
where $G_{3,0}=G_{3,0}(t)$ is the plasma concentration at time $t$ and $cl_3$, $u$ are kinetic parameters. For $G_{2,0}$ and $G_{3,0}$, closed-form solutions, described in  elsewhere~\cite{dubois2011}, were used for the calculation of plasma chemotherapy concentration.

\subsubsection{Ordinary differential equations for the intracellular module}
For each cell c at position $\bm{x}_{\mathrm{c}}$, the intracellular amount of p53, denoted by $P_{c}=P_{c
}(t)$, and the intracellular amount of VEGF, denoted by $V_{\mathrm{c}}=V_{\mathrm{c}}(t)$, are modelled by the following system of equations:
\begin{align}
\frac{dP_{\mathrm{c}} }{dt} &= k_1 - k_1^{'} \frac{K_{x_c}}{K_{p53} + K_{x_c}} P_{\mathrm{c}}, \label{eq:odep53}\\
\frac{dV_c }{d t} &= k_2 + k_2^{''} \frac{P_{\mathrm{c}}V_{\mathrm{c}}}{J_5 + V_{\mathrm{c}}} - k_2^{'} \frac{K_{x_c}} {K_{V}  + K_{x_c}}V_{\mathrm{c}} \label{eq:odevegf}
\end{align}
where $k_1$, $k_1^{'}$, $K_{p53}$, $k_2$, $k_2^{'}$, $k_2^{''}$, $J_5$ and $K_{V}$ are kinetic parameters. The function $K_{x_c} = K(\bm{x}_c,t)$ represents the oxygen concentration at cell location $x_c$ and is modelled by a PDE (described below).

Similarly, the progression though the cell cycle $\phi=\phi(t)$ of each cell c at position $\bm{x}_{\mathrm{c}}$ is modelled by the equation:
\begin{equation}
\frac{d\phi}{dt} = \frac{K_{x_c}}{T(K_{\phi} + K_{x_c})},
\label{eq:cellcycle}
\end{equation}
where $T$ and $K_{\phi}$ are scalar parameters. At start, all cancer cells are assigned a random $\phi$ value between 0 and 1. The value of the cell cycle of a newly divided cell is set to $\phi = 0$, and it increases until the first cellular update after $\phi \geq 1$. At that point the cancer cell is committed to divide and will either divide or die due to the effect of chemotherapy as described above. 
 
\subsubsection{Reaction-diffusion equations for the extravascular-extracellular module}

Each of the four drugs (Avastin $A$,  Fluorouracil $G_{1,0}$, Epirubicin $G_{2, 0}$, and Cyclophosphamide $G_{3, 0}$) is modelled as moving from the vessels (concentrations denoted by $A_1,G_{1, 0}, G_{2, 0}, G_{3, 0}$ respectively) to the extracellular-extravascular space (concentrations denoted by $A , G_1,G_2,G_3,$ respectively). These rates depend on the difference in concentrations between the vessel $U_{\rm v} = U_{\rm v} (t)$ and the EES, $U_x = U(\bm{x}, t)$, and the drugs' vessel surface permeabilities $P_{U}$, and are modelled as:
\begin{align}
R_{\mathrm{U}} =& \int_{V_0} \left(U_{\rm v}- U_x\right) P_U \, S_x \, \mathrm{d} \bm{x}
\end{align}
where the integration domain $V_0$ is the total volume of the cross-section, and $S_x = S(\bm{x})$ is the vessel surface-per-volume ratio for a vessel at $\bm{x}$.
\begin{align}
\begin{split}
R_{\mathrm{U}}=& \int_{V_0} \left(U_{\mathrm{v}} - U_x)\right)P_{\mathrm{U}} \frac{2\pi r_x h}{\Delta x^3} \sum_{\bm{x}_{\mathrm{v}} \in \bm{X}_{\mathrm{v}}}\delta(\bm{x} - \bm{x}_{\mathrm{v}}, t) \, \mathrm{d}\bm{x} \\
    =& \frac{P_{\mathrm{U}} 2\pi r_{0} h}{\Delta x^3} \int_{V_0} \left(U_{\mathrm{v}} - U_x\right)\frac{r_x}{r_{0}} \sum_{\bm{x}_{\mathrm{v}} \in \bm{X}_{\mathrm{v}}}\delta(\bm{x} - \bm{x}_{\mathrm{v}}, t) \, \mathrm{d}\bm{x}
\,,
\end{split}
\label{rate_eq_1}
\end{align}
where $r_{0}$ is the average radius of the initial vessels, $r_x = r(\bm{x}, t)$ is the radius of the vessel at $x$, and $h$ is the vessel length (which is the height of tissue section and equal to the cell size $\Delta x$). 
 Function $\delta$ is the Dirac delta function and for convenience, we introduce a short-hand for the sum of Dirac delta functions over a set of points $X_a$:
\begin{equation}
  \mathcal{N}_{a}=\mathcal{N}_{a}(\bm{x}, t) = \sum_{\bm{y} \in X_a} \delta (\bm{x} - \bm{y}, t) \label{delta_function}
\end{equation} 
This forms part of the reaction terms in \cref{eq:oxygen,eq:chemo,eq:vac_v,eq:vac_a,eq:vac_c} below. \hfill\\

For chemotherapy and Avastin, we assumed that $P_U = c_U P_{\mathrm{Gado}}$. $P_{\mathrm{Gado}}$ is the permeability of gadolinium, the contrast agent used in MRI, and $c_U \in \mathbb{R} $ is a scaling constant. Given that the initial transfer rate $K_{\mathrm{trans}}^{\mathrm{init}} = P_{\mathrm{Gado}} S^{\mathrm{init}} =  P_{\mathrm{Gado}} N_{\mathrm{v}}^{\mathrm{init}} 2\pi r_0 h / (S_{\text{MRI}} h)$, is the Tofts model permeability-surface area product of gadolinium~\cite{tofts1999}, where $N_{\mathrm{v}}^{\mathrm{init}}$ is the initial number of vessels and $S_{\text{MRI}}$ is the permeability-surface area product of a voxel in MRI, we can derive that $P_{\mathrm{U}}= c_{\mathrm{U}}K_{\mathrm{trans}}^{\mathrm{init}} S_{\text{MRI}}/(N_{\mathrm{v}}^{\mathrm{init}} 2\pi r_0)$. It can then be substituted in equation \eqref{rate_eq_1} for Avastin and the chemotherapies to give: 
\begin{align}
    R_{\mathrm{U}} = c_{\mathrm{U}} K_{\mathrm{trans}}^{\mathrm{init}} \frac{S_{\text{MRI}}}{\Delta x^2 N_{\mathrm{v}}^{\mathrm{init}} } \int_{V_0} &\frac{r_x}{r_0}\left(U_{\mathrm{v}}- U_x \right) \mathcal{N}_{v} \, \mathrm{d}\bm{x},
\end{align}


\paragraph{Oxygen}
Oxygen pressure $K = K(\bm{x}, t)$ in the EES is modeled by the following reaction-diffusion equation for all $\bm{x} \in D$ and $t > 0$:
\begin{align}   
s_K \frac{\partial K}{\partial t} = D_K \nabla^2K - \frac{\phi_K K}{K_1 + K} \mathcal{N}_{c} 
+ \frac{P_{\mathrm{K}}2\pi r_0}{\Delta x^2}(K_0 - K) \mathcal{N}_v, 
\label{eq:oxygen}
\end{align}
where $D_k$, $\phi_K$, $K_1$ and $K_0$ are scalar parameters, $X_c=\{ \bm{x} :  \mathcal{L}(\bm{x},t) \neq 0\} $ is the set of all cell locations.
$s_K \in \{0,1 \}$ is constant. When $s_K = 1$, this becomes to the time-dependent reaction-diffusion PDE, while $s_K = 0$ corresponds to the steady-state equation. The latter is used to initialise the oxygen pressure given initial locations of cells and vessels. The first term in the right-hand-side of equation (\ref{eq:oxygen}) accounts for oxygen diffusion, the second term accounts for the consumption by cells and the third term represents point sources accounting for the flow of oxygen from the blood vessels.

\paragraph{Chemotherapies}
\label{subsec:chemo}
The concentration of each chemotherapy $G_i=G_i(\bm{x}, t)$, with $i=1,2,3$ in the EES is modeled by the following reaction-diffusion equation for $\bm{x} \in D$ and $t > 0$:
\begin{align}
s_C \frac{\partial G_i}{\partial t} &= D_{G_i}\nabla^2G_i - \psi_{G_i}G_i
+ c_{\mathrm{G}} K_{\mathrm{trans}}^{\mathrm{init}} \frac{S_{\text{MRI}}}{\Delta x^2 N_{\mathrm{v}}^{\mathrm{init}} } \frac{r_x}{r_0}(G_{i,0} - G_i) \mathcal{N}_v, 
    \label{eq:chemo}
\end{align}
where $D_{G_i}$ is the diffusion constant of $G_i$, $\psi_{G_i}$ is the linear decay rate of $G_i$. $G_{i,0}$ is the concentration of chemotherapy $i$ in the blood. The first term in the right-hand-side of equation (\ref{eq:chemo}) accounts for diffusion of the concentration, the second term accounts for drug decay, the third term represent point sources accounting for the flow chemotherapy from the vessels. 
 
\paragraph{VEGF-Avastin complex}
\label{subsec:VAC}
Avastin is a VEGF-inhibitor that binds to the VEGF and produces an inactive VEGF-Avastin complex, thereby reducing the availability of active VEGF. The interactions between the Avastin, VEGF and VEGF-Avastin complex concentrations $A = A(\bm{x}, t)$, $V = V(\bm{x}, t)$ and $C = C(\bm{x}, t)$ are given by the following system of reaction-diffusion equations for $\bm{x} \in D, t > 0$:
\begin{align}
\frac{\partial V}{\partial t} &= D_V \nabla^2 V + a V_{\mathrm{c}} \mathcal{N}_c - k_a A V + k_d C - \psi_V V \label{eq:vac_v} \\
\frac{\partial A}{\partial t} &= D_A \nabla^2 A - k_a A V + k_d C - \psi_A A  + c_{\mathrm{A}} K_{\mathrm{trans}}^{\mathrm{init}} \frac{S_{\text{MRI}}}{\Delta x^2 N_{\mathrm{v}}^{\mathrm{init}} } \frac{r_x}{r_0}(A_1 - A) \mathcal{N}_v \label{eq:vac_a} \\
\frac{\partial C}{\partial t} &= D_C \nabla^2 C + k_a A V  - k_d C - \psi_C C,
\label{eq:vac_c}
\end{align}
where $D_V$, $D_A$, $D_C$, $a$, $b$, $k_a$, $k_d$, $\psi_V$, $\psi_A$, $\psi_C$ and $c_{\mathrm{A}}$ are scalar parameters and $V_{\mathrm{c}}$ is the time-dependent  intracellular concentration of VEGF described in \cref{eq:odevegf} and $A_1$ is the time-dependent concentration of Avastin intravascularly. The first term in the right-hand-side of equation (\ref{eq:vac_v}) accounts for diffusion of the VEGF concentration, the second term accounts for the release of VEGF by cells, the third and forth terms account for VEGF binding/unbinding to the VEGF inhibitor Avastin and the last term accounts for natural VEGF decay. In equation (\ref{eq:vac_a}), the first term in the right-hand-side accounts for diffusion of the Avastin concentration, the second and third terms account for Avastin binding/unbinding to VEGF, the forth term accounts for drug decay and the last term represent point sources accounting for the flow of Avastin from the vessels. Finally, the first term in the righ-hand-side of equation (\ref{eq:vac_c}) accounts for diffusion of the VEGF-Avastin complex, the second and third terms account account for Avastin binding/unbinding to VEGF and the last term represents natural decay of the complex.

\subsection{Patient-specific model initialisation and parameterisation}

All simulations run in this study were personalised by data from a specific breast cancer patient (Patient 3) from a clinical trial~\cite{lai2019}. Model initialization and parameterization are, unless differently specified, as in that publication, where clinical, histological, MRI and molecular data were use to estimate model parameters and initial values. Patient 3 is a complex patient and was chosen for exhibiting very heterogeneous perfusion conditions as observed by MRI. More precisely, MRI showed a tumor core with very low perfusion and viable cells and a tumor edge with much higher perfusion values. All parameter used for the simulations are listed in \cref{tab:parameter}
\begin{table}
    \begin{tabular}{clcc}
\hline
Parameter         & Description                                            & Value                   & Units                                                                                                         \\ \hline
$k_{\phi}$        & Oxygen concentration at half-maximal cycle speed       & 1.4                     & $\text{mmHg}$                                                                                                 \\
$k_1^{'}$         & Degration rate of p53 by oxygen                        & 0.01                    & $\text{min}^{-1}$                                                                                             \\
$K_{\text{TP53}}$ & Oxygen concentration for half-maximal TP53 degredation & 0.01                    & $\text{mmHg}$                                                                                                 \\
$k_2$             & Synthesis rate of VEGF                                 & 0.002                   & $\text{min}^{-1}$                                                                                             \\
$k_2^{'}$         & Reaction rate of p53 with VEGF                         & 0.01                    & $\text{min}^{-1}$                                                                                             \\
$J_5$             & sVEGF concentration for half-maximal sVEGF production  & 0.04                    & $\mu\text{g}~\text{mL}^{-1}$                                                                                  \\
$K_{\text{VEGF}}$ & Oxygen concentration for half-maximal VEGF degredation & 0.01                    & $\text{mmHg}$                                                                                                 \\
$D_k$             & Oxygen diffusion coefficient                           & 1.05 x 10$^{5}$         & $\mu \text{m}^2~\text{min}^{-1}$                                                                              \\
$r_k$             & Oxygen supply rate                                     & 1.88 x 10$^{4}$         & min$^{-1}$                                                                                                    \\
$K_{0}$           & Oxygen concentration in the blood                      & 20                      & $\text{mmHg}$                                                                                                 \\
$\phi_k$          & Oxygen consumption rate                                & 900                     & min$^{-1}$                                                                                                    \\
$K_1$             & Oxygen concentration for half-maximal consumption      & 2.5                     & $\text{mmHg}$                                                                                                 \\
$D_v$             & VEGF diffusion coefficient                             & 3.52 x 10$^{3}$         & $\mu \text{m}^2~\text{min}^{-1}$                                                                              \\
$a$               & VEGF secretion slope                                   & $6.66 \times  10^{-6}$  & min$^{-1}$                                                                                                    \\
$b$               & VEGF secretion intercept                               & $-1.10 \times  10^{-6}$ & $\mu \text{g}~\text{mL}^{-1}  \text{min}^{-1}$                                                                \\
$k_a$             & VEGF association rate to Avastin                       & $7.4 \times 10^{-1}$    & $\mu \text{g}^{-1}~\text{mL}~\text{min}^{-1}$                                                                 \\
$k_d$             & VEGF dissotiation rate from Avastin                    & $1.76 \times 10^{-3}$   & min$^{-1}$                                                                                                    \\
$\psi_v$          & VEGF decay rate                                        & 1.0  x 10$^{-2}$        & $\text{min}^{-1}$                                                                                             \\
$D_A$             & Avastin diffusion coefficient                          & $ 2.4 \times 10^{3}$    & $\mu \text{m}^2~\text{min}^{-1}$                                                                              \\
$\psi_C$          & Complex decay rate                                     & 1.0  x 10$^{-2}$        & $\text{min}^{-1}$                                                                                             \\
$D_{G^{i}}$       & Chemotherapies diffusion coefficient                   & 9.6 x 10$^{3}$          & $\mu \text{m}^2~\text{min}^{-1}$                                                                              \\
$\psi_{G^j}$      & Chemotherapies decay rate                              & 1.0  x 10$^{-2}$        & $\text{min}^{-1}$                                                                                             \\
$v_1$             & Avastin plasma compartment volume                      & $2.66 \times 10^3$      & $\si{\mL}$                                                                                    \\
$v_2$             & Avastin peripheral compartment volume                  & $2.76 \times 10^3$      & $\si{\mL}$                                                                                    \\
$q$               & Avastin intercompartmental clearance                   & $0.412$                 & \si{\mL \per \min}                                      \\
$cl$              & Avastin elimination clearance                          & $0.144$                 & \si{\mL \per \min}                                            \\
$v_{max}$         & Fluororacil maximal degradation rate                   & $1.75$                  & \si{\micro\gram \per \mL \per \min} \\
$k_m$             & Fluororacil half-maximal concentration                 & $27$                    & \si{\micro\gram \per \mL}                                         \\
$w_1$             & Epirubicin plasma compartment volume                   & $18 \times 10^{3}$      & \si{\mL}                                                                                    \\
$w_2$             & Epirubicin peripheral compartment volume               & $957 \times 10^{3}$     & \si{\mL}                                                                                    \\
$w_3$             & Epirubicin peripheral compartment volume               & $25 \times 10^{3}$      & \si{\mL}                                                                                    \\
$q_2$             & Epirubicin intercompartmental clearance                & $0.918 \times 10^{3}$   & \si{\micro\gram \per \mL \per \min}                                            \\
$q_3$             & Epirubicin intercompartmental clearance                & $0.25 \times 10^{3}$    & \si{\micro\gram \per \mL \per \min}                                            \\
$cl_2$            & Epirubicin elimination clearance                       & $0.983 \times 10^{3}$   & \si{\micro\gram \per \mL \per \min}                                            \\
$u$               & Cyclophosphamide plasma compartment volume             & $2430 \times 10^{3}$    & \si{\mL}                                                                                  \\
$cl_3$            & Cyclophosphamide elimination clearance                 & $3.93 \times 10^{3}$    & \si{\micro\gram \per \mL \per \min}                                           \\
$\Delta x  $      & Space interval                                         & 10                      & $\mu$m                                                                                                        \\
$\Delta t  $      & Time interval of cell cycle update                     & 30                      & min                                                                                                           \\
$\Delta v  $      & Vessel update interval                                 & 720                     & min                                                                                                           \\
$\text{Low}_{V}$  & Lower VEGF angiogenic threshold                        & 10$^{-6}$               & $\mu \text{g}~\text{mL}^{-1}$                                                                                 \\
$\alpha$          & FEC dose-response shape                                & 1                       & dimensionless                                                                                                 \\ \hline
$T_{\text{min}}$  & Minimum cell cycle duration                            & 3.74                    & days                                                                                                          \\
$k_{1}$           & Basal p53 synthesis rate                               & 0.0004                  & min$^{-1}$                                                                                                    \\
$k''_{2}$         & Maximal p53 effect in VEGF production                  & -0.0002                 & min$^{-1}$                                                                                                    \\
$d_{G^1}$         & Fluorouracil dose                                      & 600                     & mg m$^{-2}$                                                                                                   \\
$d_{G^1}$         & Epirubicin dose                                        & 100                     & mg m$^{-2}$                                                                                                   \\
$d_{G^3}$         & Cyclophosphamide dose                                  & 600                     & mg m$^{-2}$                                                                                                   \\
$d_{A}$           & Bevacizumab dose                                       & 0                       & mg m$^{-2}$                                                                                                   \\ \hline
\end{tabular}
\caption{Overview of all parameters.}
\label{tab:parameter}
\end{table}







\subsection{Numerical methods}
\label{sec:descretization}

The main algorithm for the numerical solution of the full model
equations is presented in Algorithm \ref{alg:main} below. Further specification of the numerical techniques for solving the separate subproblems are given in the text below. The numerical solver was implemented using the open source FEniCS Project finite element
library~\cite{alnaes2015}. 


\begin{algorithm}
  \caption{Numerical solution of the coupled PDE-ODE-CA model}
  \begin{algorithmic}[1]
  \item
    Initialize computational meshes $\mathcal{T}_H$ and $\mathcal{T}_h$ corresponding to the
    regular square lattice $L$ (cf.~\ref{subsec:math_model}).
  \item
    Initialize the vascular lattice representation $\mathcal{G}(\cdot,
    0)$ and cellular lattice representation $\mathcal{L}(\cdot, 0)$
    based on available patient data.
  \item
    Compute the initial oxygen concentration $K(\cdot, 0)$ by
    solving~\eqref{eq:oxygen} with $s_K = 0$.
  \item
    Define the initial condition for the cell cycle $\phi$ based on
    a uniform random distribution:
    \begin{equation*}
      \phi_{\mathbf{x}}(0) \in \mathcal{U}(0, 1) \quad \forall \, \mathbf{x} \in L, 
    \end{equation*}
    and compute the intracellular levels of $P(t)$ and $V_c(t)$ at
    initial time ($t = 0$) by solving~\eqref{eq:odep53}
    and~\eqref{eq:odevegf}, respectively, using the initial oxygen
    concentration $K(\cdot, 0)$.
  \item
    Define the initial chemotherapy concentrations by setting
    $G_j(\cdot, 0) = 0$ for $j = 1, 2, 3$. Define initial conditions
    for $V$, $A$ and $C$ by setting $V(\cdot, 0) = 0$, $A(\cdot, 0) =
    0$ and $C(\cdot, 0) = 0$.
  \item
    Compute the intravascular chemotherapy concentrations $G_{j,0}(t^i), j = 1,2,3$ and intravascular VEGF-inhibitor concentrations $A_1(t^i)$ by solving ~\eqref{eq:intraVEGFinhibitor}-\eqref{eq:intrachemo3} given
    doses $d_k, k = 1, 2, 3 \quad \forall t^i \leq T $
  \item
    Set $t^1 = \Delta t = 1$ min, $n = 1$
    \While{$t^n \leq T$}
    \State
    Compute $K(\cdot, t^n)$ by solving~\eqref{eq:oxygen} with $s_K = 1$.
    \State
    Compute $V(\cdot, t^n)$, $A(\cdot, t^n)$, $C(\cdot, t^n)$ by
    solving~\cref{eq:vac_v,eq:vac_a,eq:vac_c}.
    \State
    Compute $G^j$ for $j = 1, 2, 3$ by solving~\eqref{eq:chemo}.
    \State
    Compute $P(\cdot, t^n)$ and $V_c(\cdot, t^n)$ by
    solving~\eqref{eq:odep53} and~\eqref{eq:odevegf} and compute
    $\phi(\cdot, t^n)$ by solving~\eqref{eq:cellcycle} using
    $K(\cdot, t^n)$.
    \State
    Compute $\mathcal{L}(\cdot, t^n)$ and $\mathcal{G}(\cdot, t^n)$ based on the cellular automaton rules described in Section~\ref{subsec:CA}.
    \State
    Set $t^n = t^{n-1} + \Delta t$, $n = n + 1$.
    \EndWhile
  \end{algorithmic}
  \label{alg:main}
\end{algorithm}

\subsubsection{Numerical solution of the PDE systems}

The time-dependent, nonlinear systems of PDEs describing the evolution
of the oxygen and chemotherapy concentrations and the VEGF-Avastin
complex (\cref{eq:oxygen,eq:chemo,eq:vac_v,eq:vac_a,eq:vac_c}) are solved using the finite element method in space and finite
difference method in time. The non-linear problems are solved using the Newton-Raphson method and all linear systems are solved using iterative Krylov methods designed to scale to large scale simulations.

The computational domain $D$, representing a tissue slice, is a
rectangular region: $D=[0, (n-1)\Delta x] \times [0, (m-1)\Delta y] \subset \mathbb{R}^2$. 
The base regular lattice
defines the $n \times m$ vertices of a base triangular mesh
$\mathcal{T}$. The vertices of this mesh defines the potential
locations for the vessels and biological cells.
To avoid that the accuracy of the PDE solutions are limited by the size of the  biological cells, a finer mesh is used for the finite element discretization of the PDEs.
Specifically, we used a uniform refinement $\mathcal{T}_H$ of the base mesh with $2^2 (n - 1) \times 2^2 (m -
1)$ finite element cells. A further refined mesh $\mathcal{T}_h$ with
$2^3 (n - 1) \times 2^3 (m - 1)$ finite element cells was used for
the representation of the discrete functions i.e.~the delta function
representation of the vessels and biological cells, see Section~\ref{sec:num:cells} for more detail.

\paragraph{Solving the chemotherapy concentration equations}

We first consider the numerical solution of the system of chemotherapy
equations, \ref{eq:chemo} for $i = 1, 2, 3$. Each equation is
time-dependent but linear, and the equations are independent of each
other. We first discretize each PDE by the implicit second-order
Crank-Nicolson scheme in time. At each time $t^k$ for $k = 1, \dots,
K$, given the concentrations at the previous time $G_i^{k-1}$, we
solve for the concentrations $G_i^k$ using the finite element method
with continuous piecewise linear finite elements relative to the mesh
$\mathcal{T}_H$. The resulting linear systems of equations are
symmetric and positive definite, and were solved (with optimal
complexity) using a conjugate gradient (CG) solver with algebraic
multigrid (BoomerAMG\cite{HypreAMG}) preconditioning with a relative
solver tolerance of $10^{-8}$.

\paragraph{Solving the oxygen concentration equation}

We consider the time-dependent version of the nonlinear oxygen
concentration equation~\ref{eq:oxygen} (with $s_K = 1$). As for the
chemotherapy equations, we first discretize the PDE by the implicit
second-order Crank-Nicolson scheme in time. The resulting nonlinear
system of differential equations at each timestep $t^k$ is discretized
using continuous piecewise linear finite elements relative to the mesh
$\mathcal{T}_H$. We solve the resulting nonlinear system using a
Newton iteration, with a tolerance of $10^{-8}$. The inner loop linear
systems are symmetric and positive definite, and were solved with a CG
Krylov solver with Jacobi preconditioning, and a relative solver
tolerance of $10^{-8}$.

We also used a linear approximation of the nonlinear system that could lead to significant CPU time reduction. In particular, the non-linear term at timestep $t^k$ can be approximated by $\tfrac{K(\bm{x},t^{k})}{K_1+K(\bm{x},t^{k-1})}$, where $K(\bm{x},t^{k-1})$ denotes the solution of oxygen at time $k-1$.
The resulting linearized system of differential equations at each timestep $t^k$ is discretized using continuous piecewise linear finite elements relative to the mesh $\mathcal{T}_H$.  The resulting linear system of equation is symmetric and positive definite, and were solved (with optimal complexity) using a conjugate gradient (CG) solver with algebraic multigrid (BoomerAMG\cite{HypreAMG}) preconditioning with a relative solver tolerance of $10^{-8}$. We found that the difference between the non-linear and linear approximations to be negligible (with a difference of less than 0.03\%)

\paragraph{Solving the coupled system equations of VEGF/Avastin complex}
The system of PDEs were discretized by the implicit second-order Crank-Nicolson scheme
in time. The resulting nonlinear system of differential equations at
each timestep $t^k$ is discretized using continuous piecewise linear
finite elements relative to the base mesh $\mathcal{T}_H$. We solve
the resulting nonlinear system using Newton's method, and the
(non-symmetric) linear systems were solved with an iterative Krylov
(GMRES) solver with Jacobi preconditioning. The stopping criteria for
the Newton and Krylov solvers were set to $10^{-8}$ for both.

The coupled system of VEGF/Avastin complex can be simplified when Avastin is not administered. In this case, the PDE system reduces to a linear partial differential equation in $V$ alone. The resulting linear equation is symmetric and positive definite, and were solved (with optimal complexity) using a conjugate gradient (CG) solver with algebraic multigrid (BoomerAMG\cite{HypreAMG}) preconditioning with a relative solver tolerance of $10^{-8}$.

\subsubsection{Numerical treatment of ODEs}

The ODEs for the intravascular module and cell cycle can be solved analytically~\cite{dubois2011} and evaluated at the neccessary time points. For the coupled system of the intracellular module, solution to $P(\bm{x},t_k + dt)$ can be written in explicit form given solution $P(\bm{x},t_k)$ at $t = t_k$. For $V_c(\bm{x}, t)$, \verb+VODE+ solver with implicit Adams method from \verb+SciPy+ package was chosen for the non-stiff problem for each cell. This solver uses an adaptive time step, and and the maximum time step was set to $\SI{1}{\min}$.

\subsection{Numerical treatment of cellular automaton}
\label{sec:num:cells}

The biological cells and vessels are represented as approximate
(continuous piecewise linear) delta functions defined relative to the
refined mesh $\mathcal{T}_h$. The approximate delta function $\delta_{x_i}$ is defined such that $\delta_{x_i}$ is $\delta$ at vertex $x_i$ and zero at all other vertices, and is represented as a continuous piecewise linear function, with $\delta$ such that the integral of $\delta_{x_i}$ over its support equals 1. The refined mesh was chosen such that there is no overlap in-between different vessels and in-between different cells representation (delta) functions. The use of a refined mesh also allows for the inspection of the approximation error in the biological cell and vessel representation.

For each cancer cell at $\bm{x} = (x_i, x_j) \in \mathcal{L}$ with cell cycle
$\phi_{\bm{x}} \geq 1$, the cell is killed with probability drawn from
a cumulative Beta distribution $Beta(1,\beta), \beta > 0$. If the cell is
not killed, a new cell is placed at the lattice site $\bm{x}' \in (x_p, x_q), p=i-1,\ldots,i+1, q=j-1,\ldots,j+1 $ with the highest oxygen concentration $K$. No cell is placed if no space is available in the neighborhood, i.e. if all (Moore) neighbour sites are occupied by other cells. Next, $\phi_{\bm{x}}$ and $\phi_{\bm{x}'}$ are set to 0. The intracellular VEGF and TP53 levels $P$, $V_c$ are set to zero at both $\bm{x}$ and $\bm{x}'$ at time $t$. The sequence of updates for
proliferation-ready cells is asynchronous such that the new state of a
cell do not affect the calculation of states in neighboring cells.

\subsection{Parallelization of hybrid cellular automaton and finite element models}
\paragraph{Linking spatially parallel continuous And discrete modules}
The computational expense of the whole algorithm is dominated by the operations within each time step. Each time step consists of `continuous update' - solving a sequence of PDEs and ODEs,  and `discrete update' - computing states of cellular automaton. Parallelism in the time variable is more difficult to approach due to the inherent dependencies between the continuous and the discrete components. Instead we consider a fork-join approach as solvers within the FEniCS framework are spatially-parallel ready. They can be easily integrated with the rest of the algorithm provided the communication overhead is minimised.

The PDEs assembly and solvers were parallelized through standard domain decomposition techniques,  iterative linear solvers and preconditioners. 
The parallelization of the ODEs was trivial: they consist of spatially decoupled problems defined on each biological cell and hence can be solved independently. 
To minimize the communication between the parallel processes, we used the same domain decomposition for both $\mathcal{T}_h$ and $\mathcal{T}_H$.
The parallelization of the cellular automaton models required more attention, as they depend on the oxygen levels at the neighboring biological cells. Retrieving that information requires inter-process communication if that cell is at the boundary of the domain decomposition. 
Cells are classified into two categories depending on their location within the sub-domain. Cells located inside a sub-domain are marked as internal, while those located at the border between two or more sub-domains, shared by two, three or four processors are marked as inter-facial. To achieve this, we added a setup routine (\cref{alg:MPI}) to the algorithm in which the parallel processes collect which process owns their neighboring biological cells. These neighbor maps allowed the cellular automaton to retrieve neighboring biological cell values with minimal overhead.

\begin{algorithm}
      \begin{algorithmic}[1]
    \item MPI Initialization
    \item Make a map containing the coordinates on the mesh and the process of all eight neighbour cells for each cell.
    \item \For{each time step}
            \State Get oxygen concentration $\forall \bm{x} \in \mathcal{L} (\bm{x}) $
            \State Get and update $\phi(\bm{x})$
            \State Gather $x$ such that $\phi(\bm{x}) > 1$ 
            \State update cells according to cellular automaton rules
            \State \If{more than one processor attempt to update at the same $\bm{x}$}
                \State update $ \mathcal{L} (\bm{x}) $  from the processor with higher $\phi(\bm{x}')$ value $\forall \bm{x}'$ in the neighbour
                \State  Set automata, vegf, p53 and cell cycle value of the new cancer cell equal to the parent cancer cells
                \EndIf
           \EndFor
    \item Update map
  \end{algorithmic}

    \caption{implementation of communication between neighbour cells on different processes}
    \label{alg:MPI}
\end{algorithm}

\paragraph{Choosing a scalable parallel Random Number Generator (RNG)}
As the model is stochastic, another important aspect of the parallelisation is to have a scalable parallel RNG. To investigate and understand the simulation results, we must be able to reproduce the same scenarios and find the same confidence intervals every time we run the same stochastic experiment; when debugging such parallel stochastic application, we also need to reproduce the same result to correct any anomalous behavior. Furthermore, for applications in predicting treatment effect, it is important to produce unbiased results and claim statistical significance. Implementation of pseudo RNG also reduces the number of simulations needed for claiming statistical significance in cross-comparison between different treatment regimens since the simulation would be the same until the point of divergence induced by different regimens.

Pseudo RNG in high-performance computing (HPC) applications, particularly in our HCA model, requires the following criteria. It cannot download, or store the hundreds of thousands of numbers needed to reproduce the experiments due to constraints of the server. The random number sequence assigned to each cell or vessel must not depend on the number of processors, i.e. each cell or vessel should autonomously obtain either its own random sequence or its own subsequence of a global sequence. The pseudo RNG must also come with good statistical properties, approximating as close as possible a truly random sequence. The RNG should also be memory-efficient as sequences of the RNG will be the length of any given simulation, and the size of the RNG grows at the rate of the total number of vertices. We have therefore chosen permuted congruential RNG \cite{oneill:pcg2014} which partitions the main sequence of the generator into subsequences for the sake of memory-efficiency, good statistical properties, small state and a more than sufficient period.

\section{Results}

\subsection{Small-scale simulations}\label{sec}

\begin{figure}
    \centering
    \includegraphics[width=\textwidth]{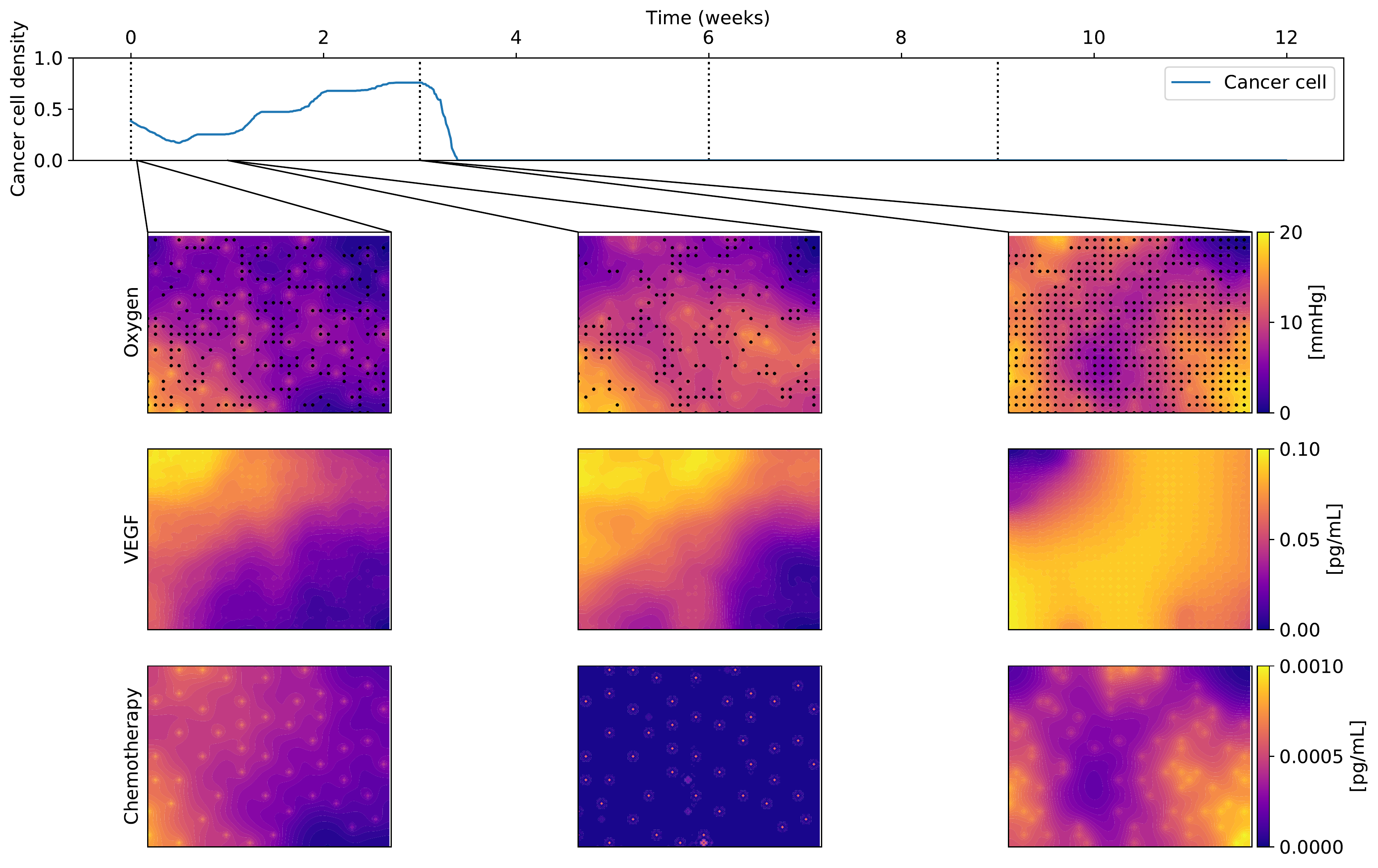}
    \caption{
    Simulation of a small tissue cross section of size $\SI{200}{\micro\meter} \times \SI{300}{\micro\meter} $. The top panel depicts the evolution of the cancer cell density over the course of 12 weeks of treatment. Chemotherapy shots are given at weeks 0, 3, 6 and 9 and are illustrated by a vertical dotted line in the figure. The lower panels depict the spatial distribution of oxygen, VEGF and chemotherapy at weeks 0, 1, and 3 in the tissue section. The presence of cancer cells is shown in black dots in the oxygen panel. Since the vessels are point sources of those molecules, areas with high blood vessel density can be perceived in the oxygen and chemotherapy panels.}
    \label{fig:p3_original}
\end{figure}

We first run personalised simulations of a patient analyzed previously with a previous version of the model and using a non-parallel solver\cite{lai2019}. These simulations correspond to a $\SI{200}{\micron} \times \SI{300}{\micron}$ tissue section, a size that allowed us to use exact initial cell positions available from histopathological slides. We use here the same initial conditions and previously published model parameters~\cite{lai2019} (see \cref{tab:parameter}).  As both the model and the solver are now extended, we do not expect to get the same solutions as in previous work. We tested different perfusion conditions estimated from MRI images and confirmed that the treatment outcome after 12 weeks strongly depends on the perfusion conditions used. In \Cref{fig:p3_original}, we show a simulation representing a high perfusion condition estimated at the outer edge of the tumor. In the simulation the initial cell density is reduced by half after the first chemotherapy shot at week 0. Chemotherapy is efficiently distributed in space but it is washed out after a few days. As approximately half of the cells are not dividing while the chemotherapy is available, they are not killed and the number of cells grows significantly before the second chemotherapy shot at week 3.  After week 3 instead, all cancer cells are killed as they divide when enough chemotherapy is available. Comparing with previous simulations, the final outcome is matched but the transients are slightly different. We attribute it mostly to solving time-dependent PDEs instead of the steady-state solver considered previously.  This difference in solvers adds a factor of seven to the run time on a single core.

\subsection{Solver performance}

To evaluate the runtime and scalability of the complete numerical
solver, we consider a simulation of a $\SI{1}{\mm^2}$ tissue slice
($D = [0, 1] \times [0, 1]$ mm$^2$), with $n = m = 100$.  Initial cell and vascular densities were assumed to be 0.4 and 0.12 respectively. These settings reflect a real patient simulation scenario.

\subsubsection{Serial runtime profile}

\begin{figure}[ht]
    \centering
    \begin{subfigure}[b]{0.48\textwidth}
        \centering
        \includegraphics[width=\linewidth]{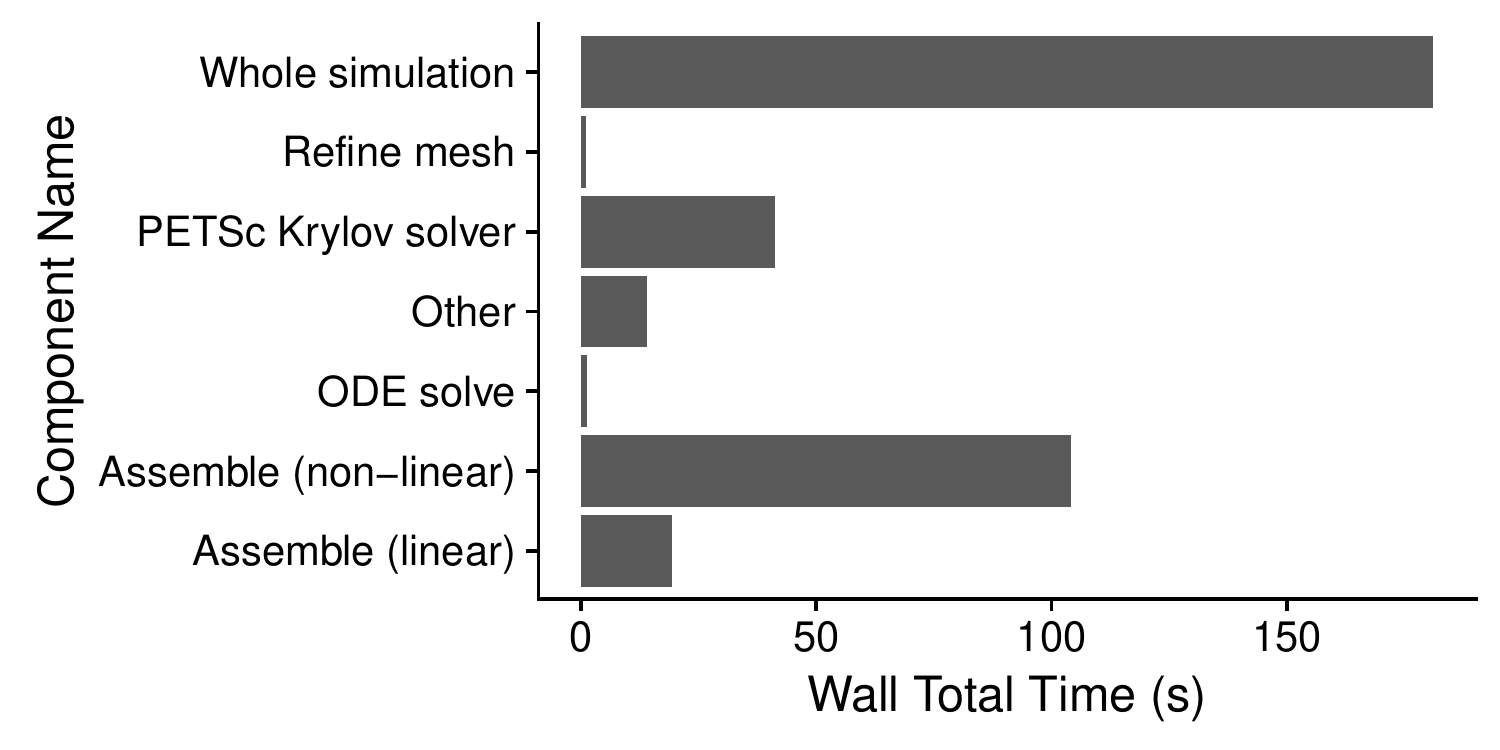}
        \caption{Intra-update timing of the solver grouped by processes}
        \label{fig:intra_timing_a}
    \end{subfigure}%
    ~ 
    \begin{subfigure}[b]{0.48\textwidth}
        \centering
        \includegraphics[width=\linewidth]{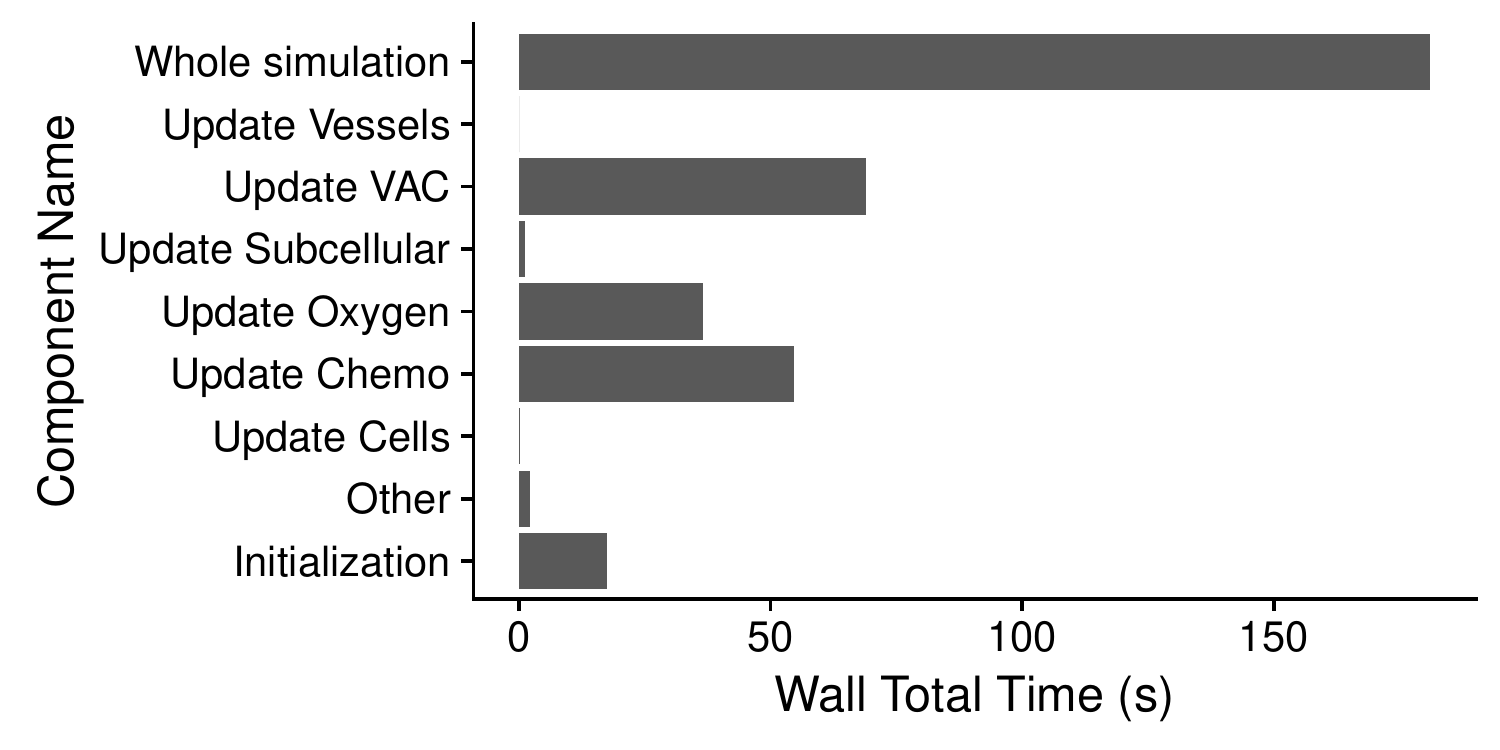}
        \caption{Intra-update timing of the solver grouped by modules of the model}  
        \label{fig:intra_timing_b}
    \end{subfigure}
    \caption{Intra-update timing of the model with $\Delta t = \SI{1}{min}$ solved for $t=\SI{30}{min}$.}
    \label{fig:intra_timing}
\end{figure}

\begin{figure}[ht]
    \centering
    \begin{subfigure}[b]{0.48\textwidth}
        \centering
        \includegraphics[width=\linewidth]{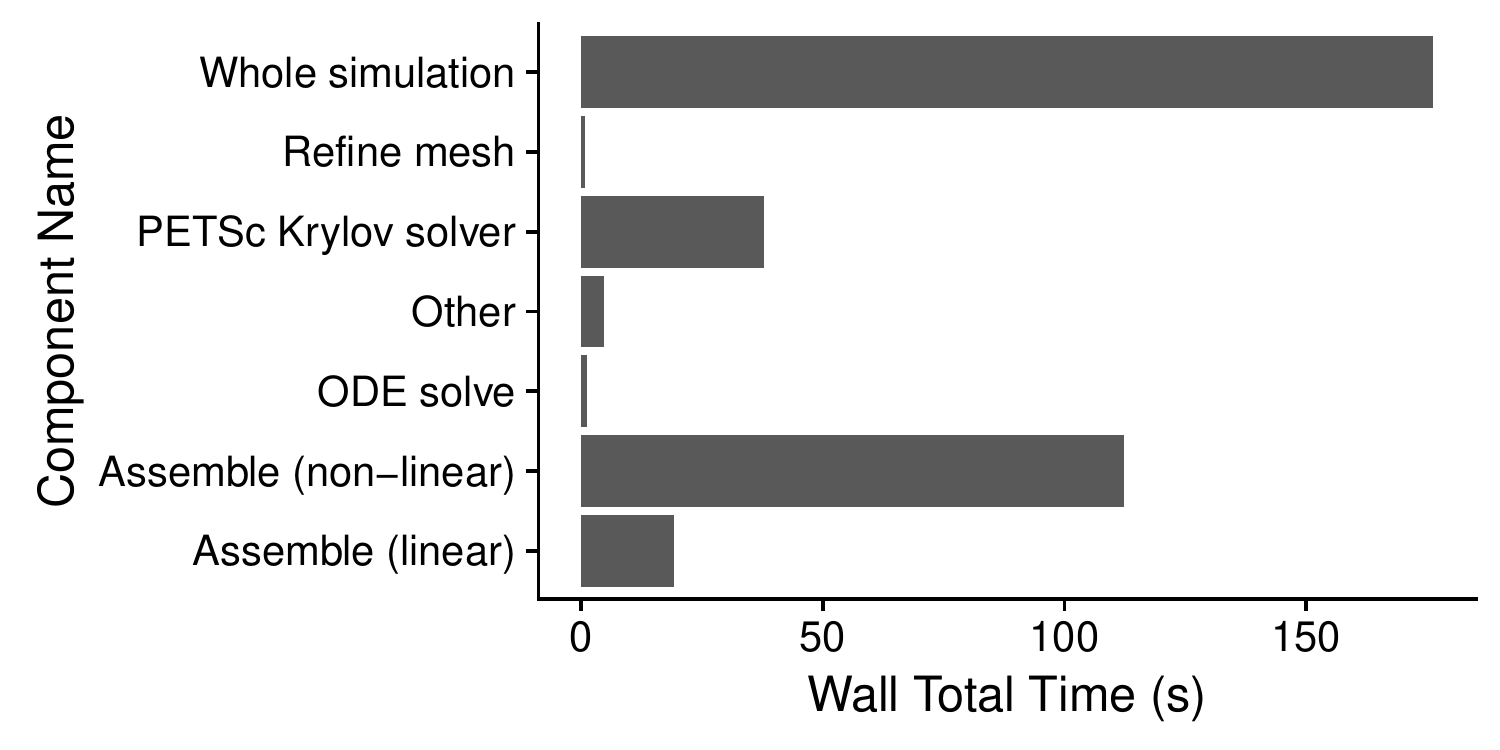}
        \caption{Average inter-update timing of the solver grouped by processes}
        \label{fig:inter_timing_a}
    \end{subfigure}%
    ~ 
    \begin{subfigure}[b]{0.48\textwidth}
        \centering
        \includegraphics[width=\linewidth]{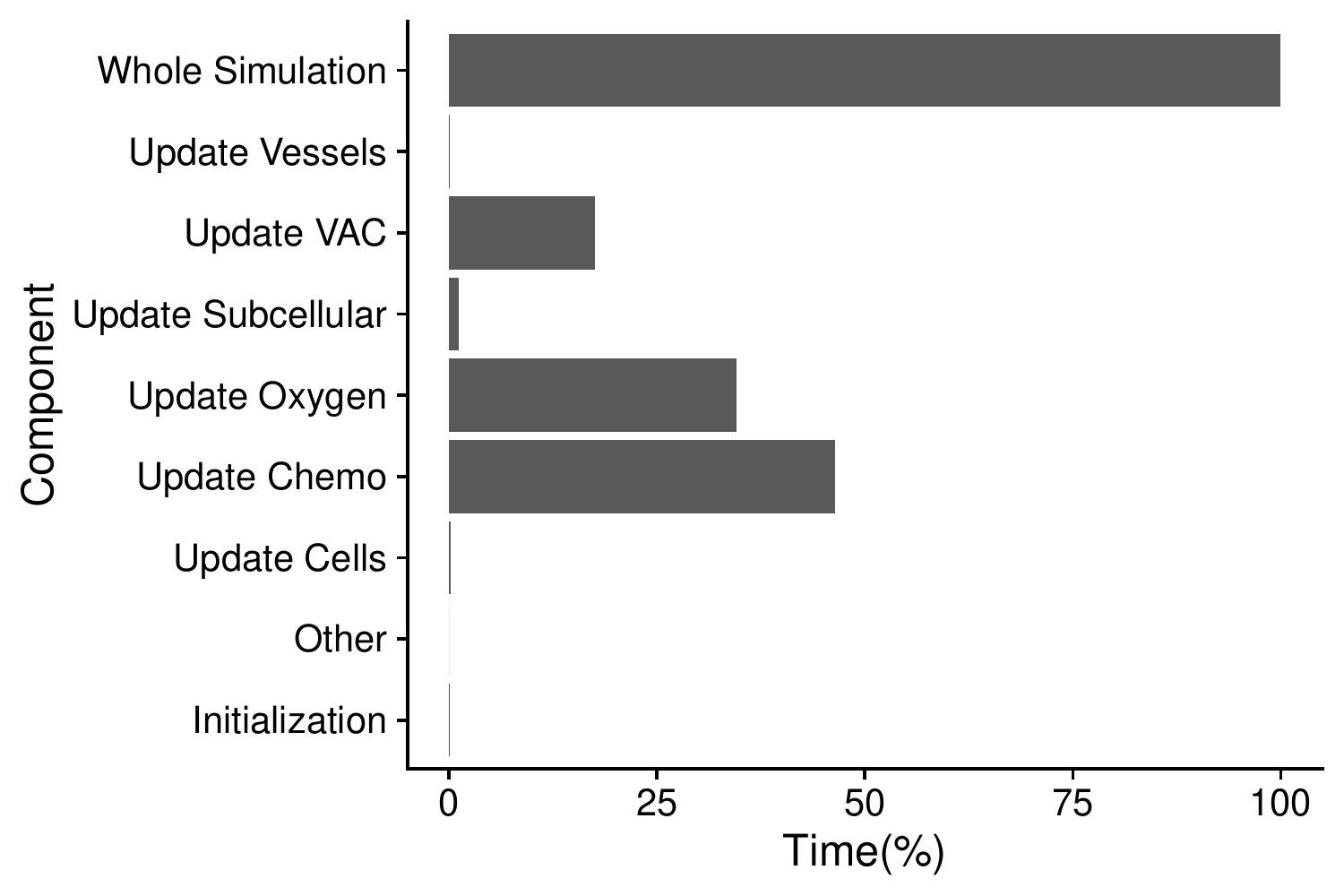}
        \caption{Average inter-update timing of the solver grouped by modules of the model}
        \label{fig:inter_timing_b}
    \end{subfigure}
    \caption{Average inter-update timing of the model with $\Delta t = \SI{30}{min}$ solved over $t=\SI{120}{min}$.}
    \label{fig:inter_timing}
\end{figure}

The average intra- and inter-update timing of the solver was first
carried out on a single CPU core. The total timings, broken down by
computational processes and by model modules, are shown in Figures~\ref{fig:intra_timing}
and \ref{fig:inter_timing}.

We observe that the assembly of the discrete PDE operators dominates
the runtime, specifically the time-dependent assembly of the
non-linear PDEs (\Cref{fig:intra_timing_a}). As expected, the solution
of the linear systems (PETSc Krylov solver) is also a substantial
component. The other components and in particular the solution of the
ODEs represent a nearly negligible cost. Overall we found similar cost
distribution comparing intra- and inter-update, indicating that
communication during discrete updates were efficient.

The breakdown by modules shed some further insights on the cost of
assembly: the solution of both VAC and Oxygen modules are obtained by
solving non-linear PDEs, and the system is re-assembled for each
$\Delta t$. To reduce the runtime of these modules, an explicit
splitting scheme could be used instead (corresponding to a single step
of the Newton iteration) at the cost of numerical accuracy. Time spent
on chemotherapy module comes second as there are three types of
chemotherapies to be solved separately. Initialization includes mesh
initialization as well as function initialization for each
module. This is done only once at the beginning of the whole
simulation, and is therefore expected to become negligible for longer simulations.

\subsubsection{Parallel scalability}

We also investigated the parallel (weak) scalability of the assembly
and solution of the PDEs. Weak scalability can be examined for a
series of problem sizes by assigning a constant problem size to each
processing element -- typically for an increasing amount of
processor cores. Perfect weak scaling is achieved if the run time stays
constant while the workload is increased in direct proportion to the
number of processors. The weak scaling factor was defined as
\begin{align}
    {\rm WSF} = \frac{t_N}{t_1} ,
\end{align}
where $t_1$ and $t_N$ is the runtime when using 1 and $N$ processors,
respectively, and was calculated for the assembly and solve times
separately.

All weak scaling experiments were conducted on Saga,
the high performance computing facility placed at NTNU in Trondheim,
composed of Intel Xeon-Gold 6138 running at 2.0 GHz. Each Saga compute node consists of 40 physical compute cores with 192 GiB memory each, giving approximately 4 GiB memory per physical core. We used FEniCS version 2019.1.0 built with an Intel compiler. These scalability tests were run on two compute nodes totalling 80 physical cores, and with exclusive access
to minimize communication overhead.
For the sake of comparison, the end time of all temporal updates of
PDEs in the model was set to $t = 30$ min, while cell and vessel
configurations were initialized randomly.
The parallel runtime and scalability of the chemotherapy equations are
presented in~\Cref{tab:chemo_timing}. We observe a nearly optimal WSF
($\leq 1.3 $) for the finite element assembly up to $80$ cores. For the
linear solver, we observe a moderate increase in the WSF up to $2$ for
$16$ cores and $2.5$ for $80$ cores. The number of iterations of the
Krylov solver stay constant however. 
\begin{table}
\begin{center}
    \begin{tabular}{cccccccc}
\toprule
$N_{\rm proc}$ & $N_{\rm it}$ & $N_{\rm dofs}$ & $\bar{N}_{\rm dof}$ & $t_A$ & WSE$_A$ & $t_S$ & WSE$_S$ \\
\midrule
1 & 5 & 1 & 159197 & 2.9177 & 1 & 6.2043 & 1 \\
8 & 6 & 8 & 1279157 & 3.1415 & 1.0767 & 9.8944 & 1.5948 \\
16 & 6 & 16 & 2556797 & 3.543 & 1.2143 & 12.911 & 2.081 \\
24 & 6 & 24 & 3837677 & 3.6139 & 1.2386 & 14.752 & 2.3777 \\
32 & 6 & 32 & 5121165 & 3.8919 & 1.3339 & 15.248 & 2.4577 \\
40 & 6 & 40 & 6405957 & 3.7023 & 1.2689 & 15.502 & 2.4986 \\
80 & 6 & 80 & 12809237 & 3.7708 & 1.2924 & 15.784 & 2.544 \\
\bottomrule
\end{tabular}

\end{center}
\caption{Timing of temporal chemotherapy update ($t$ = 30 min). $N_{\rm
    proc}$ is the number of processors, $N_{\rm it}$ is the number of
  Krylov iterations, $N_{\rm dofs}$ is the total number of degrees
  of freedom, $\bar{N}_{\rm dof}$ is the average number of degrees of
  freedom per process, $t_{A}$ is assembly runtime, WSE$_A$ is the
  weak scaling efficiency for the assembly, $t_{S}$ is the linear
  solver runtime, and WSE$_{S}$ the linear solver weak scaling
  efficiency. $\Delta t$ was kept constant during the update and the
  results shown are the minimum of 5 repeated runs of the updates with
  random initialization of cell and vessel
  configurations. Intravascular concentration was kept constant during
  the simulation.}
\label{tab:chemo_timing}
\end{table}

For the solution of the oxygen concentration equation, the parallel scalability timings are shown in~\Cref{tab:oxygen_timing}. We note that the
solution time is around four times that of the chemotherapy equations and thus a dominant contribution. The finite element assembly scales well up to $80$ cores (WSE$_A \leq 1.4$) for both non-linear and the linearized solver. However, by linearizing the oxygen equation, the weak scalability of the linear solver is improved from a WSF of up to 9 in 80 cores to be comparable to the results for the chemotherapy equations (WSE less than 2.24).
We also investigated the weak parallel scalability of the VAC equations. The results (data not shown) were highly comparable to the scalability of the oxygen concentration solver.
\begin{table}[ht]
\begin{subtable}{\textwidth}
    \centering
    \begin{tabular}{cccccccc}
\toprule
$N_{\rm proc}$ & $N_{\rm it}$ & $N_{\rm dofs}$ & $\bar{N}_{\rm dof}$ & $t_A$ & WSE$_A$ & $t_S$ & WSE$_S$ \\
\midrule
1 & 12 & 159197 & 159197 & 13.112 & 1 & 44.012 & 1 \\
8 & 12 & 1279157 & 159894 & 15.211 & 1.1601 & 82.138 & 1.8663 \\
16 & 12 & 2556797 & 159799 & 15.528 & 1.1843 & 164.76 & 3.7435 \\
24 & 11 & 3837677 & 159903 & 15.802 & 1.2052 & 277.56 & 6.3065 \\
32 & 11 & 5121165 & 160036 & 15.753 & 1.2014 & 296.43 & 6.7352 \\
40 & 10 & 6405957 & 160036 & 18.601 & 1.419 & 322.40 & 7.3253 \\
80 & 10 & 6405957 & 160036 & 18.485 & 1.410 & 399.94 & 9.0871 \\
\bottomrule
\end{tabular}
    \caption{Timing of temporal oxygen update ($t$ = 30 min).}
    \label{tab:oxygen_timing_a}
\end{subtable}
\hfill
\begin{subtable}{\textwidth}
    \centering
    \begin{tabular}{cccccccc}
\toprule
$N_{\rm proc}$ & $N_{\rm it}$ & $N_{\rm dofs}$ & $\bar{N}_{\rm dof}$ & $t_A$ & WSE$_A$ & $t_S$ & WSE$_S$ \\
\midrule
1 & 1 & 159197 & 159197 & 2.6657 & 1 & 3.0532 & 1 \\
8 & 8 & 1279157 & 159894 & 3.0377 & 1.1396 & 3.9054 & 1.2791 \\
16 & 16 & 2556797 & 159799 & 3.3315 & 1.2498 & 5.6732 & 1.8581 \\
24 & 24 & 3837677 & 159903 & 3.5559 & 1.3339 & 6.2294 & 2.0403 \\
32 & 32 & 5121165 & 160036 & 3.5187 & 1.32 & 6.6993 & 2.1942 \\
40 & 40 & 6405957 & 160148 & 3.7548 & 1.4086 & 6.6725 & 2.1854 \\
80 & 80 & 12809237 & 160115 & 3.5985 & 1.3499 & 6.8429 & 2.2412 \\
\bottomrule
\end{tabular}
    \caption{Timing of temporal linearized oxygen update ($t$ = 30 min).}
    \label{tab:oxygen_timing_b}
\end{subtable}
\caption{Timing of temporal oxygen update ($t$ = 30 min).  $N_{\rm
    proc}$ is the number of processors, $N_{\rm it}$ is the number of
  Krylov iterations, $N_{\rm dofs}$ is the total number of degrees of
  freedom, $\bar{N}_{\rm dof}$ is the average number of degrees of
  freedom per process, $t_{A}$ is assembly runtime, WSE$_A$ is the
  weak scaling efficiency for the assembly, $t_{S}$ is the linear
  solver runtime, and WSE$_{S}$ the linear solver weak scaling
  efficiency. $\Delta t$ was kept constant during the update and the
  results shown are the minimum time of 5 repeated runs of the updates
  with random initialization of cell and vessel configurations.}
\label{tab:oxygen_timing}
\end{table}

\subsection{Simulating a cross-section of a MRI voxel of tissue}

The spatial resolution of the available MRI data is defined by the imaging voxels of size \SI{1}{\mm^3}. Thus, important biological information, such as the perfusion of the tumor is only given per voxel of tissue. As an intermediate step to run multi-voxel simulations, here we demonstrate the ability of our parallel solver to run a grid-size equivalent to a cross-section of a MRI voxel.

\begin{figure}
    \centering
    \includegraphics[width=\textwidth]{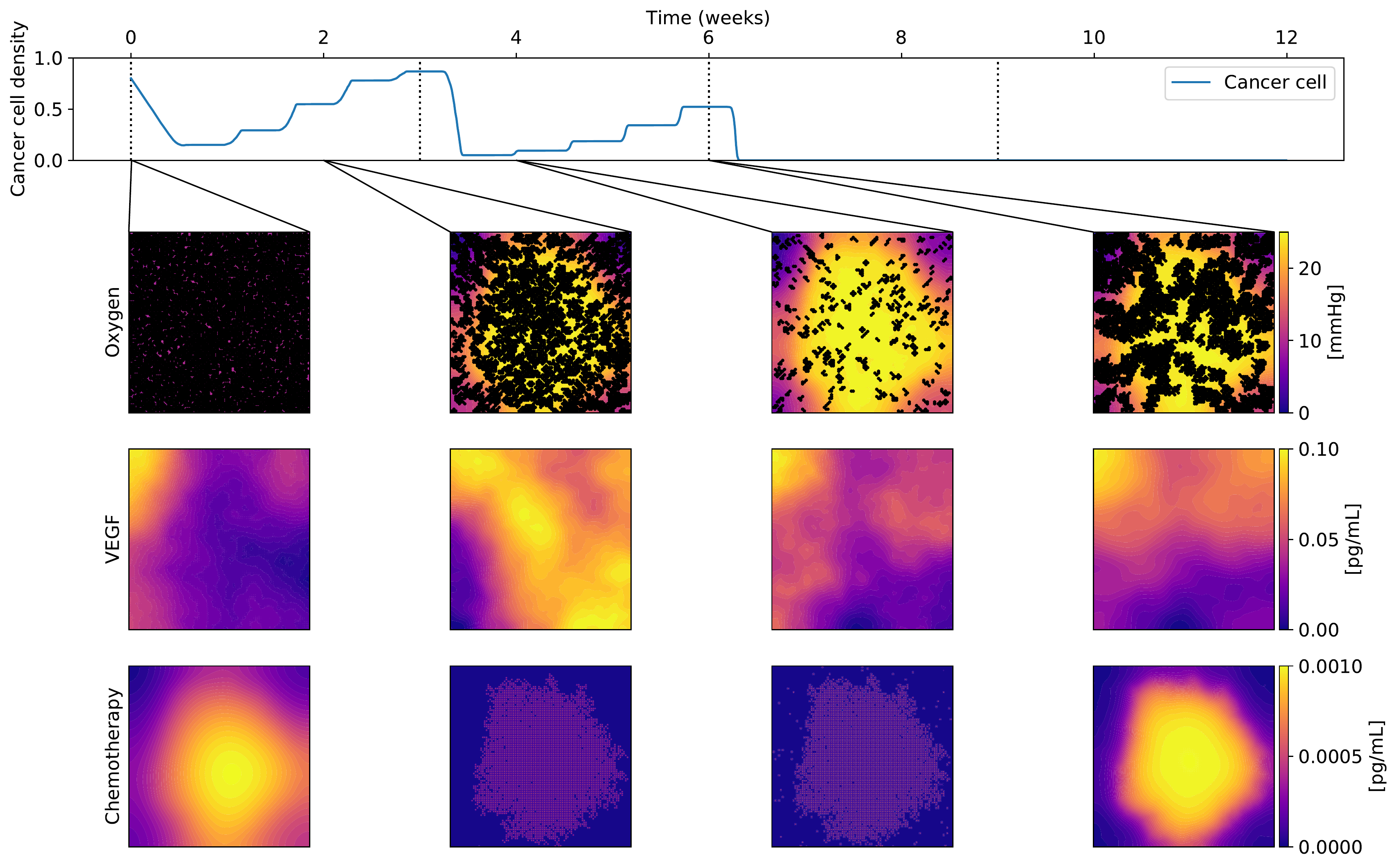}
    \caption{Simulation of a patient biopsy of size $\SI{1}{\mm} \times \SI{1}{\mm} $, equivalent to the size of a MRI voxel. The top panel depicts the evolution of the cancer cell density over the course of 12 weeks of treatment. Chemotherapy shots are given at weeks 0, 3, 6 and 9 and are illustrated by a vertical dotted line in the figure. The lower panels depict the spatial distribution of oxygen, VEGF and chemotherapy at weeks 0, 2, 4 and 6 in the tissue section. The presence of cancer cells is shown in black dots in the oxygen panel. Since the vessels are point sources of those molecules, areas with high blood vessel density can be perceived in the oxygen and chemotherapy panels.}
    \label{fig:p3_mri}
\end{figure}

\Cref{fig:p3_mri} illustrates the cancer cell trajectory under 12 weeks of chemotherapy treatment. Since digitalized biopsy data detailing the exact distribution of cancer and stroma cells are available only for the small tumor section of size $\SI{200}{\micron} \times \SI{300}{\micron}$, we used the same cell density to approximate the number of cancer and stroma cells in the larger one, but they were randomly distributed in the grid to initialize the simulation. We again use the same model parameters detailed in \cref{tab:parameter}. Comparing Figure~\ref{fig:p3_mri} to Figure~\ref{fig:p3_original}, we observe similar patterns. Immediately following the initial administration of chemotherapy, the number of cancer cells starts to decrease, reaching a minima about 2.5 days later. At that point, the chemotherapy treatment loses its effect and the cancer cells start to grow back again. This pattern is repeated after the second chemotherapy shot at 3 weeks, while the third shot at 6 weeks kills all existing cancer cells. The panels at weeks 0, 2, 4 and 6 in \cref{fig:p3_mri} show the spatial distribution of cells, oxygen, VEGF, blood vessels and chemotherapy in the simulated tumor cross-section. As expected, the tissue oxygen concentration increases and VEGF concentration decreases when the number of cancer cells decreases. Due to a good tissue perfusion with a large number of vessels and high permeability, chemotherapy reaches the whole tissue at effective concentrations. The total runtime of this 12-week simulation on Saga took approximately 12 hours with 80 cores.

\subsection{Simulating a cross-section of a digital biopsy sample}

To ultimately demonstrate the impact of heterogeneous perfusion conditions on the dynamics of cell growth and killing, we constructed a mesh of size equivalent to $2 \times 10$ MRI voxels. This matches with the size of the core needle biopsy taken at the beginning of the treatment in the clinical trial and we will refer to it as a digital biopsy sample. Perfusion related parameters were applied in each of the twenty simulated voxels of a digital biopsy sample as follows. First, the MRI slice with the widest tumor diameter is selected. This image reveals a difference in perfusion conditions between the tumor edge and the core. Starting from the boundary between the core and the edge of the tumor, 5 voxels were chosen radially inwards and outwards respectively, totalling 10 voxels. Another 10 voxels were chosen in the same fashion from their immediate neighbour, resulting in a simulated biopsy of $2 \times 10$ voxels. Estimated perfusion parameters, specifically $k_{trans}$ and $v_p$, from the extended Tofts model were then calculated individually to each voxel. 
Since exact distribution of vessels, cancer and stroma cells are currently available only for small tumor sections, we used $v_p$ and $v_c = 1 - (v_p + v_e)$ (where $v_e$ is the estimated fractional extravascular extracellular space volume), from applying the extended Tofts model to the MRI data as estimates of the vessel and cell density respectively. Vessels and cells were then randomly distributed accordingly within each voxel in the grid ($100 \times 100$ grid points) to initialize the simulation.

\begin{figure}
    \centering
    \includegraphics[width=\textwidth]{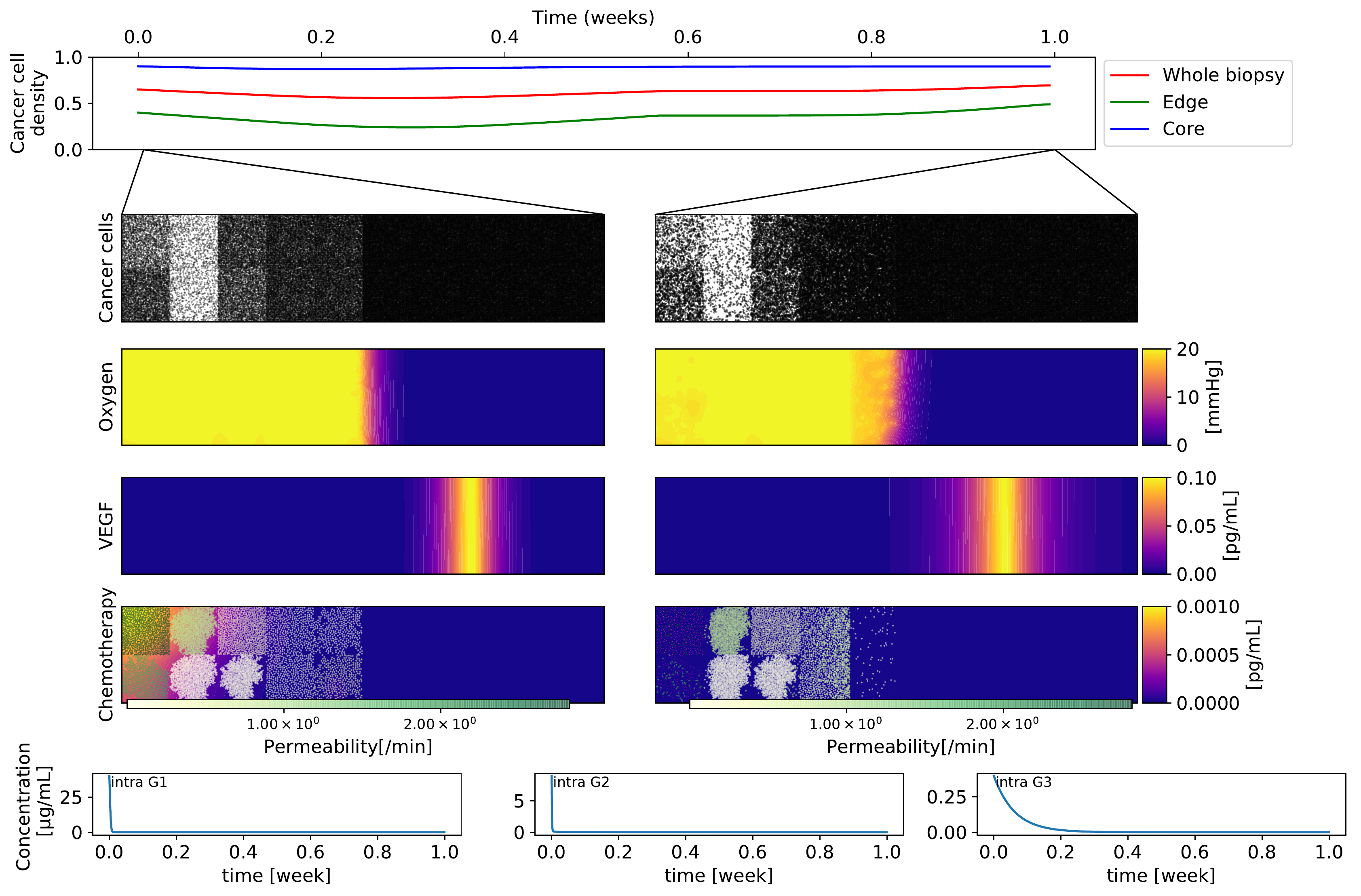}
    \caption{Simulation of a digital biopsy sample of size $\SI{2}{\mm} \times \SI{1}{\cm} $ from week 0 to week 1. The top panel depicts the evolution of the cancer cell density of the whole, the edge and the core of the simulated biopsy over the course of 1 week of treatment. This is followed by spatial comparative view of cancer cells distribution and three extracellular variables, from top to bottom, oxygen, VEGF, and chemotherapy between week 0 (left panels) and week 1 (right panels). In the first row, locations of simulated cancer cells were marked in black, while locations of simulated vessels were overlaid in color to represent their permeability in chemotherapy. Intravascular concentration of the three chemotherapies between week 0 and week 1 are shown at the bottom.}
    \label{fig:p3_full}
\end{figure}

\Cref{fig:p3_full} shows one week evolution of one digital biopsy sample under therapy. We can clearly see the complex effect of the heterogeneous perfusion condition illustrated by the heterogeneity in the oxygen concentration. 
As chemotherapy reaches the tissue, active cancer cells are killed at different rates in various areas of the simulated biopsy. In areas with high perfusion, cells are killed more rapidly, while in those ares with low perfusion, chemotherapy cannot be delivered sufficiently, and more cancer cells survive. In areas with no vessels, cell growth is stalled and bounded by the capacity of the computational grid and the oxygen. In areas with lower initial cell density, cells were dividing albeit at a lower rate. In addition, a combined effect in areas between high and low perfusion is observed. Drug delivery to areas with low perfusion reaches neighboring areas of high perfusion by diffusion instead of via the vasculature. Moreover, dense cell populations from low perfusion areas can grow into areas with high perfusion where more cancer cells were killed by the chemotherapy. Total runtime of the simulation was on average 18 hours with 80 cores.

To assess the validity of our model, we extracted 10 biopsy samples using the same method described above.  We use data from week 0 of each digital biopsy to initialize the model and differences in cell density between week 0 and week 1 are evaluated between the simulation and the MRI data for each voxel within the 10 biopsy samples. Given that all digital biopsies were collected from the same tumor, a mixed effect model was used to verify if there is any statistical difference between simulated and actual densities. We found no significant difference, thus proving that the numerical simulations can reproduce the observed drug outcome in large, clinically-relevant heterogeneous tumor portions. 

\begin{figure}[ht]
    \centering
        \includegraphics[width=\textwidth]{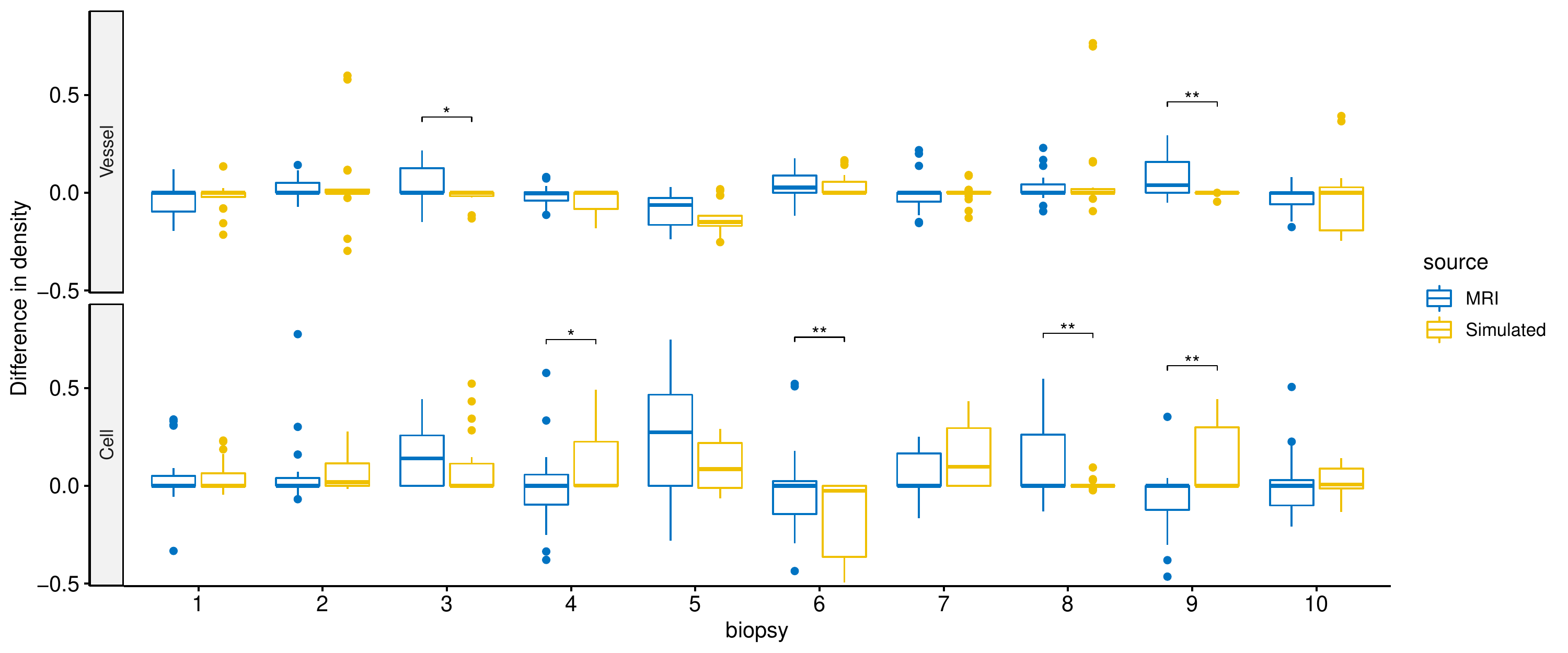}
    \caption{Differences in cell densities between weeks 0 and 1 are depicted as box-plots. Simulated and actual cell densities from MRI were seen in blue and yellow, respectively. The x-axis shows the comparison of ten biopsy slices. The first box-plot from the left compares the simulation shown in \cref{fig:p3_full} to MRI data. Paired Wilcoxon signed rank test were performed between simulated and actual cell densities for each biopsy. Asterisks were placed above the boxes of comparisons of p-values less than 0.05. * $p<0.05$ ** $p<0.01$ }
    \label{fig:validate_MRI}
\end{figure}

\begin{table}[ht]
\centering
\begin{tabular}{|l|l|l|l|l|} 
\toprule
Cells                                                & Estimate & Std. Error & t-value & p-value  \\ 
\hline
(Intercept)                                                  &  0.04917292 & 0.02014897 & 2.4404685 & 0.0151  \\ 
\hline
Simulated density compared to actual density & -0.01138260 & 0.02021387 & -0.5631087  & 0.5737   \\
\bottomrule
\end{tabular}

\begin{tabular}{|l|l|l|l|l|} 
\toprule
Vessels                                                & Estimate & Std. Error & t-value & p-value  \\ 
\hline
(Intercept)                                                  &   0.003049063 & 0.01643971 & 0.1854693 & 0.8530  \\ 
\hline
Simulated density compared to actual density
&  0.002905701 & 0.01412006 & 0.2057853 & 0.8371   \\
\bottomrule
\end{tabular}

\caption{Results from the fitted mixed effect model. Intercept and the source of the cell/vessel density (simulated or actual) were considered as the fixed effects; the location of each MRI voxel and the biopsy sample ID was considered as nested random effects to reflect the variation in biopsy samples in either simulated or actual vessel and cell densities. The regression coefficients estimates, standard error as well as the significance of the two fixed effects were shown in the table. }
\end{table}

\section{Discussion}
\label{sec:discussion}
In this work, we present a multiscale HCA model for personalized breast cancer growth under the effect of therapy, together with a parallel numerical algorithm aimed at large scale, clinically-relevant simulation sizes. 
We discovered that the PDE solvers are the most computationally intensive component of the system, with non-linear coupled equation updates dominating the runtime. 
In terms of solver performance, the assembly of discrete PDE operators scales almost linearly, while a moderate increase in weak scaling efficiency was observed in linear solver runtime.
We simulated tumor dynamics and drug therapies of tumor cross-sections in three different system sizes: a small tumor portion, an MRI voxel of tissue, and a full biopsy sample consisting of 20 MRI voxels.
Our results show that the parallel implementation of our model can account for tumor heterogeneity at a clinically-relevant system size while correctly predicting treatment outcome.


Previously, multiple cancer modelling studies have used HCA models that couple CA with ODEs and PDEs, see for instance \cite{alarcon2005, gerlee2007, perfahl2011, powathil2012, lai2019}. However, they typically considered small 2D or 3D tumor portions and account for only small number of cells. Multi-scale models representing clinically relevant tumor portions, like the one considered in this study, require parallel computing, specially if they include many PDEs describing the tumor microenvironment or very large ODE systems for cell signalling. A number of computational frameworks for simulating multi-scale models with parallel computing capabilities are available~\cite{metzcar2019review}, including Morpheus~\cite{starruss2014}, CompuCell3D~\cite{swat2012}, PhysiCell~\cite{ghaffarizadeh2018}, CellSys~\cite{hoehme2010}, Chaste~\cite{mirams2013}, Biocellion~\cite{kang2014} or Timothy~\cite{cytowski2014}. Each of these tools have their own specific features regarding modelling formalism, implementation, usability and performance. Morpheus and CompuCell3D use cellular Potts model as cell-based model formalism while CellSys, PhysiCell, Biocellion and Timothy use other off-lattice/cell centered approaches.  Chaste has more flexibility and allows the user to choose their own cell-based formalism, including CA like in our case. In terms of performance, several simulation frameworks allow multithreading via OpenMP, including Morpheus, CompuCell3D, PhysiCell and CellSys. Biocellion and Timothy, instead, were designed for intense parallelisation between nodes and allow domain decomposition techniques as our solver. These tools can be used for simulating large numbers of PDEs in large domains and up to billions of biological cells in high performance supercomputers. In order to run personalised simulation of breast cancer therapy considered in our previous work~\cite{lai2019} but considering clinically relevant tumor portions, we have taken this later approach. Specifically, we build on the domain decomposition capabilities of FEniCS for solving PDEs and efficiently link the stochastic CA models with the continuous PDE and ODE models. To the best of our knowledge, this is the first time that the popular finite element framework FEniCS is extended for simulating a multi-scale HCA model.




In terms of limitations, for the initialization of cells, we turned to the use of estimated cell density from MRI rather than real cell distribution captured from the core biopsy staining. This is due to lack accurate identification of cancer and stroma cells in the used dataset. In this way, we do not consider local differences in cell density in each voxel and thus simulated treatment outcomes might differ from the actual outcomes in some cases.  In future work, we plan to digitalise the pathological sample and incorporate cell identification. Considering other cell types and their interactions would be also possible. We did not account for it due to inability of the present clinical data to inform them on a patient-specific basis. Each MRI voxel contains a large number of vessels, with varying surface permeability and vascular flow, both properties that are relevant to drug delivery and response.
In this study, a voxel was treated as a region of uniform vascular flow and vascular permeability. Improving on this will necessitate MRI data with a higher spatial resolution than that available in the clinical trial under consideration here.
Steady-state models were used in our previous implementation.
They are often understood as a simplification of the time evolution ones, with the assumption that the time-dependent solution stabilizes around the steady-state as $t \rightarrow \infty$. 
In our study, where temporal accuracy is essential, this assumption is invalid for slow perfusion scenarios. However, since it is beyond the reach of our work, we have only put in a limited amount of effort to tune the multigrid preconditioners for this problem in order to improve simulation run-time of 1 week of actual treatment, which is roughly 6 hours using 80 cores in cluster).


The possibility of simulating clinically relevant pieces of breast tumors under therapy is an important step towards the development of \textit{in silico}-guided clinical trials for personalized cancer medicine. In such trials, the idea is to compare a standard approved therapy with personalized drug schedules optimized using computer simulations. To achieve this using complex models like the one presented here, improved statistical inference methods are needed, that allow successful estimation of patient-specific parameters. We have made the first steps in that direction using Approximate Bayesian computation~\cite{henri-inprep,kohnluque2020, laiphd2019}, a very computationally intensive technique that certainly takes benefit of the scalable solver presented in this study.  

\section{Acknowledgments}
The simulations were performed on resources provided by UNINETT Sigma2 - the National Infrastructure for High Performance Computing and Data Storage in Norway.  This project received funding from the European
Union's Horizon 2020 Research and Innovation Programme under Grant Agreement No. 847912. The project received funding from the UiO: Life
Science initiative through the convergence environment grant PerCaThe. X.L, A.K.L and A.F. were supported by the center for research-based-innovation  BigInsight. We acknowledge funding from the Research Council of Norway with project number 294916. The authors also acknowledge the Centre for Digital Life Norway for supporting the partner projects PerCaThe, PINpOINT and BigInsight. M.E.R has received funding from the European Research Council (ERC) under the European Union's Horizon 2020 research and innovation programme under grant agreement 714892.

\bibliography{wileyNJD-AMA}%

\end{document}